\documentclass[11pt,twoside]{article}
\usepackage[T1]{fontenc}
\usepackage{lmodern}
\usepackage[a4paper,left=2cm,right=2cm,top=2cm,bottom=2cm,headsep=0.7cm]{geometry}
\usepackage{graphicx}
\usepackage{amsmath,amssymb,amsfonts}
\usepackage{amsthm}
\usepackage{booktabs}
\usepackage{array}
\usepackage{xcolor}
\usepackage[font=small,labelfont=sc,labelsep=period,justification=justified,singlelinecheck=true]{caption}
\usepackage{microtype}
\usepackage{titlesec}
\usepackage{lettrine}
\usepackage{fancyhdr}
\usepackage[hidelinks]{hyperref}
\graphicspath{{Figures/}}
\renewcommand{\S}{\ensuremath{\mathsection}}  

\renewcommand{\thesection}{\arabic{section}}
\renewcommand{\thesubsection}{\thesection.\arabic{subsection}}
\numberwithin{equation}{section}
\titleformat{\section}{\centering\normalsize\bfseries}{\S\,\thesection.}{0.5em}{\MakeUppercase}
\titleformat{\subsection}{\centering\normalsize\bfseries}{\thesubsection.}{0.5em}{}
\titleformat{\subsubsection}[runin]{\normalsize\itshape}{\thesubsubsection.}{0.5em}{}[.]
\titleformat{\paragraph}[runin]{\normalsize\itshape}{}{0pt}{}
\titlespacing*{\section}{0pt}{16pt plus 3pt minus 2pt}{8pt plus 2pt}
\titlespacing*{\subsection}{0pt}{12pt plus 2pt minus 2pt}{6pt plus 1pt}
\titlespacing*{\paragraph}{\parindent}{4pt plus 1pt}{0.5em}

\fancypagestyle{plain}{\fancyhf{}\fancyfoot[C]{\footnotesize\thepage}}

\newtheoremstyle{classic}{6pt}{6pt}{\normalfont}{\parindent}{\scshape}{.}{0.5em}{\thmname{#1}\thmnumber{ #2}\thmnote{ (#3)}}

\theoremstyle{classic}

\newcommand{\dd}{\mathrm{d}}
\newcommand{\e}{\mathrm{e}}
\newcommand{\erfc}{\operatorname{erfc}}
\newcommand{\lind}{\ell_{\mathrm{ind}}}
\newcommand{\lrec}{\ell_{\mathrm{rec}}}
\newcommand{\MCJ}{M_{\mathrm{CJ}}}
\newcommand{\TCJ}{T_{\mathrm{CJ}}}
\newcommand{\pCJ}{p_{\mathrm{CJ}}}
\newcommand{\uCJ}{u_{\mathrm{CJ}}}
\newcommand{\cCJ}{c_{\mathrm{CJ}}}

\begin{document}

\thispagestyle{plain}
\begin{center}
  {\Large\bfseries The final stages of a Chapman--Jouguet detonation: support by its own products\par}
  \vspace{14pt}
  {\normalsize\textsc{Prabakaran Rajamanickam}$^{1}$ \ and \ \textsc{Adam D. Weiss}$^{2}$\par}
  \vspace{6pt}
  {\footnotesize\itshape $^{1}$Department of Mathematics \& Statistics, University of Strathclyde, Glasgow G1 1XQ, United Kingdom\par}
  {\footnotesize\itshape $^{2}$ATA Engineering, Inc., San Diego 92128, USA\par}
  \vspace{3pt}
  {\small prabakaran.rajamanickam@strath.ac.uk\par}
  \vspace{6pt}
  {\small November 18, 2026\par}
  \vspace{14pt}
  \begin{minipage}{0.86\textwidth}\small\noindent\hspace{1.5em}A planar Chapman--Jouguet detonation that runs out of fuel must become a shock. We show that, when the chemical energy decreases in the direction of propagation, the wave does not pass through the local Chapman--Jouguet states. The products of the richer mixture are at a higher pressure than the local Chapman--Jouguet state, so the wave is overdriven, it loses its sonic point, and forward characteristics carry the state of the rich products to the shock. The rarefaction behind the wave removes this support only if the distance travelled since initiation is small compared with the thickness of the fuel gradient, which is rarely the case. The wave speed instead approaches an envelope given by a Riemann problem between the rich products and the local mixture, ending at the Mach number of the shock transmitted into inert gas. For strong detonations this Mach number tends to a fraction $C$ of the Chapman--Jouguet value, with $0.621<C<0.625$ for every ratio of specific heats. The mechanism is purely gas-dynamic. Shock-fitted computations with a two-step chain-branching mechanism, for gradients 1 to 1000 induction lengths thick, confirm that no sonic point forms and that the induction zone never decouples.\par
  \vspace{6pt}
  \end{minipage}\par
  \vspace{12pt}
  \rule{0.2\textwidth}{0.4pt}\par
\end{center}
\vspace{6pt}

\section{Introduction}\label{sec:intro}

\lettrine[lines=2,lhang=0.05,nindent=0.1em]{D}{etonations} in open space are necessarily finite, because the supply of reactant is finite. A detonation travelling through a combustible cloud must reach the edge of the cloud, and there it must give up being a detonation. Beyond the edge it continues as a shock, now unsupported by heat release, which weakens as it propagates into the surrounding inert gas. The purpose of this paper is to describe how this transition takes place.

The edge of a cloud is never perfectly sharp. A mixture prepared some time before ignition has diffused into its surroundings, so the detonation meets a gradual decrease of fuel concentration over some distance $a$. The obvious first guess is that, when $a$ is large compared with the length $\lind$ of the induction zone, the wave adjusts gradually. It would then pass through a sequence of states, each close to the Chapman--Jouguet (CJ) detonation of the local composition, until the local CJ wave can no longer exist. We shall show that this guess is generally wrong, and that the reason lies not in the chemistry but in the gas behind the wave.

A CJ detonation is followed by a rarefaction that brings the products from the CJ state to rest. It was described independently by Zeldovich~\cite{Zeldovich1940} and Taylor~\cite{Taylor1950}, and we call it the Zeldovich--Taylor wave. The CJ wave is isolated from this rarefaction by its sonic point: at the end of the reaction zone the gas leaves the shock at exactly the local speed of sound, so no disturbance from behind can reach the shock. Now suppose the chemical energy of the mixture \emph{decreases} in the direction of propagation. The CJ pressure of the leaner mixture ahead is lower than the pressure of the products already formed from richer mixture. Those products push on the wave, which becomes overdriven relative to the local CJ state. The sonic point disappears, and the products, which carry the memory of the richer mixture, communicate with the shock. They can be prevented from doing so only if the Zeldovich--Taylor wave lowers their pressure quickly enough. This requires its length $L_0$, the distance travelled by the detonation since initiation, to be small compared with $a$, which is seldom the case. The asymmetry deserves emphasis. When the chemical energy \emph{increases} in the direction of propagation, the new products have the higher pressure, expand freely backwards, and no such support arises.

The chemical model needs care. With a one-step Arrhenius reaction, quenching requires an \emph{ad hoc} cut-off temperature, and the large activation energies that make asymptotic analysis sharp make the planar CJ wave violently unstable~\cite{Erpenbeck1964,LeeStewart1990}. Both difficulties are removed by chain-branching kinetics of the type introduced by Zeldovich~\cite{Zeldovich1948} and Li\~{n}\'{a}n~\cite{Linan1971}, in the form used by Dold and co-workers~\cite{Dold2002,Dold2007} (see also~\cite{LinanWilliams1993}), which we call the Zeldovich--Li\~{n}\'{a}n--Dold (ZLD) mechanism. A fuel F and a chain carrier Z react through a thermally neutral branching step, $\mathrm{F}+\mathrm{Z}\to2\mathrm{Z}$, with a large activation energy, and through an exothermic recombination step, $\mathrm{Z}+\mathrm{M}\to\mathrm{P}+\mathrm{M}$, whose rate is independent of temperature. Branching outpaces recombination only above a crossover temperature, and this is the physical origin of the cut-off. Li\~{n}\'{a}n~\cite{Linan1971} showed that the structure of the reaction zone depends on how fast recombination is relative to branching, and distinguished the regimes of slow, intermediate and fast recombination. His classification has organised much of the subsequent theory of chain-branching flames~\cite{Joulin1985,SanchezWilliams2014}, including the recent analysis of near-limit flames by Lee \& Kim~\cite{LeeKim2024}, the non-modal stability of planar flames~\cite{Napieralski2027}, and propagation below the flammability limit~\cite{Carmona2026}. For detonations, Short and co-workers~\cite{ShortQuirk1997,ShortSharpe2003} showed that chain-branching waves are stabilised when the heat-release zone is long compared with the induction zone, and that the crossover temperature defines a detonability limit~\cite{ShortQuirk1997}; see also~\cite{Ng2005,SharpeMaflahi2006,Sanchez2001}.

The closest related problem is the decay of a plane overdriven detonation when its supporting piston is suddenly stopped, analysed by Clavin \& Denet~\cite{ClavinDenet2018}. There the coupling between the unsteady inner structure of the detonation and the centred rarefaction in the burnt gas governs the relaxation to the CJ regime. The present problem is, in a sense, its complement: the support persists while the mixture ahead weakens. The decay of a shock overtaken by a rarefaction has a long history, going back to Chandrasekhar~\cite{Chandrasekhar1943} and Friedrichs~\cite{Friedrichs1948}. Clavin \emph{et al.}~\cite{Clavin2021} showed that the dynamics of the sonic point in the rarefaction behind the reaction zone controls the critical behaviour of spherical detonations. Huete \emph{et al.} analysed the interaction of strong, overdriven detonations with density and vorticity perturbations, first treating the detonation as a discontinuity~\cite{Huete2013} and then resolving its finite-rate heat-release zone~\cite{Huete2014}. In those studies the overdrive is imposed from behind and the perturbations are small. Here the overdrive is measured relative to the local CJ state of a mixture that changes by an amount of order unity, and it is supplied by the products of the detonation itself. Analytical theories of detonation dynamics, in particular in the limit of small heat release, are reviewed by Clavin \& Williams~\cite{ClavinWilliams2012} and Clavin \& Searby~\cite{ClavinSearby2016}; see also~\cite{ClavinHe1996,ClavinWilliams2002,ClavinWilliams2009,Clavin2004,Clavin2019,ClavinDenet2020}.

The complementary problem, in which a gradient of initial temperature decides whether a detonation forms at all, was posed by Zeldovich \emph{et al.}~\cite{Zeldovich1970}. Detonation propagation through composition gradients has been studied numerically, mainly in two dimensions with detailed or one-step chemistry~\cite{IshiiKojima2007,Kessler2012}. Cuadra \emph{et al.}~\cite{Cuadra2020} developed a linear theory for planar detonations, treated as discontinuities, crossing small random fluctuations of the fuel mass fraction about a uniform mean. They found that such fluctuations raise the mean propagation speed at second order for weakly overdriven detonations, and that the linear thin-detonation description loses validity as the CJ condition is approached, where the acoustic perturbations grow without bound. The present problem differs in three respects. The change of composition is large and monotonic, and the fuel vanishes altogether. The incoming wave is a self-sustained CJ wave, the case that the linear theory cannot reach. And the reaction zone is resolved, because the outcome depends on how its length compares with the thickness of the layer.
Here we isolate the one-dimensional mechanism, describe it by characteristics and by a Riemann problem, and compare the description with shock-fitted computations. Multidimensional effects, which enter when the layer is comparable with the cell size or with the radius of the front, are excluded.

The paper is arranged as follows. The problem is formulated in \S\ref{sec:formulation}, and the planar CJ wave with ZLD kinetics is described in \S\ref{sec:CJ}. Section~\ref{sec:support} contains the theory of the supported wave: its characteristic structure, the Riemann envelope, the sharp-layer limit and the regimes of the problem. The unsteady computations are described in \S\ref{sec:unsteady}, and \S\ref{sec:conclusions} contains concluding remarks. The appendices give the structure of the induction zone, the formulae for the Zeldovich--Taylor wave, and the local-CJ description that would apply if $L_0\ll a$.

\section{Formulation}\label{sec:formulation}

\subsection{The mixing layer}

Consider a quiescent mixture at temperature $T_0$, pressure $p_0$ and density $\rho_0$, with sound speed $c_0=(\gamma p_0/\rho_0)^{1/2}$, in which the fuel mass fraction $Y_F$ varies with position $x$ as
\begin{equation}
Y_F = \frac{\bar Y}{2}\,\erfc\!\left(\frac{x}{a}\right), \qquad Y_F(-\infty)=\bar Y, \quad Y_F(+\infty)=0 .
\label{eq:layer}
\end{equation}
Here $\bar Y$ is the fuel mass fraction of the rich mixture and $a$ measures the thickness of the mixing layer. The error-function profile is the one produced by diffusion from an initially sharp interface. A planar CJ detonation appropriate to the fuel content $\bar Y$ was initiated at a closed end, and travels in the direction of increasing $x$. At time $t=0$ its shock is at $x=x_0$, upstream of the layer, and a distance $L_0$ from the closed end. We assume that the fuel, the chain carrier, the products and the inert diluent have the same molecular weight and the same constant specific heats $c_p$ and $c_v$, with ratio $\gamma=c_p/c_v$. The mixing layer therefore contains no gradient of density or sound speed: the wave sees only a gradient of chemical energy. Transport processes are neglected, as is usual in detonation theory.

\subsection{Chemical kinetics}

The ZLD mechanism consists of three steps,
\begin{equation}
\begin{aligned}
&\text{0 (initiation):}\quad && \mathrm{F}\to\mathrm{Z}, && \omega_I = \rho\,\varepsilon\, k_B(T)\, Y_F,\\
&\text{I (branching):}\quad && \mathrm{F}+\mathrm{Z}\to 2\mathrm{Z}, && \omega_B = \rho\, k_B(T)\, Y_F\, Y_Z/\bar Y,\\
&\text{II (recombination):}\quad && \mathrm{Z}+\mathrm{M}\to\mathrm{P}+\mathrm{M}, && \omega_C = \rho\, k_C\, Y_Z .
\end{aligned}
\label{eq:steps}
\end{equation}
Here $Y_Z$ is the mass fraction of the chain carrier, $\rho$ the density, $T$ the temperature, and $\omega_I$, $\omega_B$, $\omega_C$ are the masses converted by each step per unit volume per unit time. The rate coefficients are
\begin{equation}
k_B(T) = A_B\,\e^{-E/RT}, \qquad k_C = \text{constant},
\label{eq:rates}
\end{equation}
where $A_B$ is a pre-exponential factor, $E$ the activation energy of branching and $R$ the universal gas constant. The fuel and the chain carrier have the same enthalpy of formation, so steps 0 and I release no heat. The chemical energy, $q$ per unit mass of chain carrier, is released by the recombination step II. Its rate is linear in $Y_Z$ and nearly independent of temperature, as for $\mathrm{H}+\mathrm{O}_2+\mathrm{M}\to\mathrm{HO}_2+\mathrm{M}$ in hydrogen--air~\cite{SanchezWilliams2014,fernandez2019one,carpio2020near}. In the reduced four-step mechanism for methane--air~\cite{Peters1985,PetersWilliams1987}, the same division appears: the branching step proceeds at the rate of $\mathrm{H}+\mathrm{O}_2\to\mathrm{OH}+\mathrm{O}$, which has a large activation energy, while the global recombination step, written $\mathrm{H}+\mathrm{H}+\mathrm{M}\to\mathrm{H}_2+\mathrm{M}$, proceeds at the combined rate of three elementary reactions with no activation energy, the leading one being $\mathrm{H}+\mathrm{O}_2+\mathrm{M}\to\mathrm{HO}_2+\mathrm{M}$. As in earlier model studies of chain-branching detonations~\cite{ShortQuirk1997,Ng2005}, the dependence of the rates on density has been suppressed. The initiation step serves only to start the chain. We give it the temperature dependence of branching and a small relative strength $\varepsilon\ll1$, so that its only leading-order effect is the factor $\ln(1/\varepsilon)$ in the induction length (\S\ref{sec:CJ}).

Chain carriers are produced by branching at the rate $\omega_B$ and removed by recombination at the rate $\omega_C$. In undiluted reactant, $Y_F=\bar Y$, the two rates balance when $k_B(T)=k_C$. The temperature $T_c$ at which this happens is the \emph{crossover temperature}:
\begin{equation}
k_B(T_c) = k_C, \qquad\text{so that}\qquad \frac{k_B(T)}{k_C} = \exp\left[\frac{E}{R}\left(\frac{1}{T_c}-\frac{1}{T}\right)\right].
\label{eq:Tc}
\end{equation}
Below $T_c$ a chain carrier is more likely to recombine than to branch, and the chain cannot run away. In a mixture whose fuel content is a fraction $y_0=Y_F/\bar Y$ of the full value, the balance is $y_0k_B(T)=k_C$. Using (\ref{eq:Tc}) and solving for $T$ gives the crossover temperature of that mixture,
\begin{equation}
T_c(y_0) = \frac{T_c}{1 + (RT_c/E)\ln y_0} \quad \text{in which} \quad T_c=T_c(1).
\label{eq:Tcy}
\end{equation}
Since $\ln y_0<0$, the crossover temperature \emph{rises} as the fuel is depleted.

\subsection{Governing equations}\label{sec:governing}

Let $u$ be the gas velocity and $p$ the pressure. Conservation of mass and momentum, the first law of thermodynamics for a fluid particle, and the conservation of each species give
\begin{gather}
\frac{\partial\rho}{\partial t}+\frac{\partial(\rho u)}{\partial x}=0, \qquad
\rho\frac{Du}{Dt} = -\frac{\partial p}{\partial x},
\label{eq:dim1}\\
\rho c_v\frac{DT}{Dt} = -p\frac{\partial u}{\partial x} + q\,\omega_C,
\label{eq:dim2}\\
\rho\frac{DY_F}{Dt} = -\omega_I-\omega_B, \qquad
\rho\frac{DY_Z}{Dt} = \omega_I+\omega_B-\omega_C,
\label{eq:dim3}
\end{gather}
where $D/Dt=\partial/\partial t+u\,\partial/\partial x$ is the derivative following a fluid particle. The gas is ideal, $p=\rho R_gT$, with $R_g=c_p-c_v$. Equation (\ref{eq:dim2}) states that the internal energy of a particle increases through compression work and through the heat released by recombination.

We measure lengths with $a$, velocities with $c_0$, time with $a/c_0$, temperature with $T_0$, density with $\rho_0$ and pressure with $\rho_0c_0^2$. Mass fractions are measured with $\bar Y$, ... so that 
\begin{equation}
    y=\frac{Y_F}{\bar Y}, \qquad z = \frac{Y_Z}{\bar Y}, \qquad y_p = \frac{Y_P}{\bar Y},
\end{equation}
where $Y_P$ is the mass fraction of products. Steps 0--II conserve the total mass of F, Z and P, so $y+z+y_p$ is constant along each particle path. It equals the fuel content that the particle had before it was shocked. With these units the equation of state becomes $\gamma p=\rho T$, because $\rho_0c_0^2=\gamma\rho_0R_gT_0$. Equations (\ref{eq:dim1})--(\ref{eq:dim3}) become
\begin{gather}
\frac{\partial\rho}{\partial t}+\frac{\partial(\rho u)}{\partial x}=0, \qquad
\frac{Du}{Dt} = -\frac{1}{\rho}\frac{\partial p}{\partial x}, \qquad \gamma p = \rho T,
\label{eq:nd1}\\
\frac{DT}{Dt} + (\gamma-1)T\frac{\partial u}{\partial x} = \frac{2\gamma Q}{\gamma+1}\,\Pi\, z,
\label{eq:nd2}\\
\frac{Dy}{Dt} = -\Pi\,\Omega, \qquad
\frac{Dz}{Dt} = \Pi\,(\Omega - z),
\label{eq:nd3}
\end{gather}
where all variables are now dimensionless, $T_c$ denotes the crossover temperature divided by $T_0$, and
\begin{equation}
\Omega = y\,(z+\varepsilon)\,\exp\left[\beta\left(\frac{1}{T_c}-\frac{1}{T}\right)\right]
\label{eq:Omega}
\end{equation}
is the combined rate of initiation and branching measured in units of $k_C$. In obtaining (\ref{eq:nd2}) we divided (\ref{eq:dim2}) by $\rho c_v$ and used $p/(\rho c_v)=(\gamma-1)T$. The coefficient of the source term is $\gamma q\bar Y/(c_pT_0)$, which equals $2\gamma Q/(\gamma+1)$ in terms of the parameter $Q$ defined below. The parameters are
\begin{equation}
Q = \frac{\gamma+1}{2}\frac{q\bar Y}{c_pT_0}, \qquad \beta = \frac{E}{RT_0}, \qquad T_c, \qquad \varepsilon, \qquad \Pi = \frac{a\,k_C}{c_0}, \qquad \frac{L_0}{a}.
\label{eq:params}
\end{equation}
The heat-release parameter $Q$ fixes the CJ Mach number of the incoming wave~\cite{Williams1985},
\begin{equation}
\MCJ = \frac{D_{\mathrm{CJ}}}{c_0} = \sqrt{Q}+\sqrt{Q+1}, \qquad\text{equivalently}\qquad Q = \frac{(\MCJ^2-1)^2}{4\MCJ^2},
\label{eq:MCJ}
\end{equation}
where $D_{\mathrm{CJ}}$ is the CJ speed. The Damk\"{o}hler number $\Pi$ compares the acoustic time across the layer, $a/c_0$, with the recombination time $1/k_C$. In \S\ref{sec:CJ} we find that the induction length of the incoming wave is $\lind=\hat\ell_1\,c_0/k_C$, where $\hat\ell_1$ is a number fixed by $\gamma$, $Q$, $\beta$, $T_c$ and $\varepsilon$. It is then more natural to measure the layer thickness with $\lind$ and use
\begin{equation}
\Lambda = \frac{a}{\lind} = \frac{\Pi}{\hat\ell_1}
\label{eq:Lambda}
\end{equation}
in place of $\Pi$. Unless otherwise stated we take
\begin{equation}
\gamma = 1.4, \qquad \MCJ = 5\ (Q=5.76), \qquad \beta = 50, \qquad T_c = 3.3, \qquad \varepsilon = 10^{-4},
\label{eq:values}
\end{equation}
and regard $\Lambda$ and $L_0/a$ as the parameters to be varied. For these values the shock of the incoming wave heats the gas to $5.8$, well above $T_c$. A shock heats the ambient gas to $T_c$ only if its Mach number exceeds about $3.5$.

\subsection{Initial and boundary conditions}

Ahead of the shock the gas is undisturbed: $u=0$, $\rho=T=1$, $y=\tfrac12\erfc x$ and $z=0$. The ambient gas does react, but at a rate smaller than behind the shock by a factor of about $\exp[-\beta(1-1/5.8)]\approx10^{-18}$, and this is ignored. At $t=0$ the flow behind the shock is the steady CJ structure of \S\ref{sec:CJ} followed by the Zeldovich--Taylor wave. The boundary conditions used in the computations, including the treatment of the shock, are described in \S\ref{sec:unsteady}.

\section{Chapman--Jouguet waves with chain-branching kinetics in a uniform mixture}\label{sec:CJ}

\subsection{Steady equations}

Consider first a steady CJ wave travelling at constant speed into a uniform mixture of fuel content $y_0\le1$. Its Mach number is $M=\MCJ(y_0)=\sqrt{Qy_0}+\sqrt{Qy_0+1}$, from (\ref{eq:MCJ}) with $Q$ replaced by $Qy_0$, since the available chemical energy is proportional to the fuel content. In a frame moving with the shock the flow is steady. Since the mixture is uniform, the layer thickness $a$ plays no role in this section, and we measure lengths instead in units of $c_0/k_C$, the distance travelled at the speed of sound in one recombination time. All quantities then depend only on the distance behind the shock, which we denote by $s$; in terms of the dimensionless variables of \S\ref{sec:formulation}, $s=\Pi\,(x_s-x)$, where $x_s$ is the shock position. Let $w=M-u>0$ be the speed of the gas relative to the shock.

Because the flow is steady in the moving frame, $\partial/\partial t=-M\,\partial/\partial x$, and the particle derivative becomes
\begin{equation}
\frac{D}{Dt} = (u-M)\frac{\partial}{\partial x} = -w\frac{\partial}{\partial x} = \Pi\,w\,\frac{\dd}{\dd s}.
\label{eq:Dsteady}
\end{equation}
Similarly, each conservation law in (\ref{eq:nd1})--(\ref{eq:nd2}) becomes the statement that a flux is independent of $s$. Equating each flux to its value ahead of the shock, where $\rho=1$, $w=M$, $p=1/\gamma$ and $T=1$, gives
\begin{equation}
\rho w = M, \qquad p+\rho w^2 = \frac1\gamma + M^2, \qquad \frac{T}{\gamma-1}+\frac{w^2}{2} = \frac{1}{\gamma-1}+\frac{M^2}{2}+\frac{2Q}{\gamma^2-1}\,y_p .
\label{eq:RH}
\end{equation}
These are the conservation of mass, momentum and total enthalpy across the wave. The last term in the energy relation is the heat released so far, where $y_p=y_0-y-z$ by the conservation noted in \S\ref{sec:governing}, and $q\bar Y/c_0^2=2Q/(\gamma^2-1)$ by (\ref{eq:params}) and $c_0^2=(\gamma-1)c_pT_0$. Since heat is released only by recombination, the thermodynamic state depends on $y_p$ alone.

Eliminating $\rho$ and $p$ from (\ref{eq:RH}) with $\gamma p=\rho T$ gives a quadratic equation for $w$. For a wave of arbitrary Mach number $M$ its subsonic root is
\begin{equation}
w = \frac{1+\gamma M^2 - \sqrt{(M^2-1)^2-4Qy_p M^2}}{(\gamma+1)M}, \qquad
T = \frac{\gamma w}{M}\left(\frac1\gamma+M^2-Mw\right),
\label{eq:wgeneral}
\end{equation}
the second relation following from the momentum equation and the equation of state. For the CJ wave, $(M^2-1)^2=4Qy_0M^2$ by (\ref{eq:MCJ}), and the square root becomes $(M^2-1)\sqrt{1-y_p/y_0}$:
\begin{equation}
w = \frac{1+\gamma M^2 - (M^2-1)\sqrt{1-y_p/y_0}}{(\gamma+1)M}.
\label{eq:wT}
\end{equation}
At the shock, where $y_p=0$, this gives the Neumann state~\cite{LandauLifshitz1987},
\begin{equation}
w_n = \frac{(\gamma-1)M^2+2}{(\gamma+1)M}, \qquad
T_n = \frac{[2\gamma M^2-(\gamma-1)][(\gamma-1)M^2+2]}{(\gamma+1)^2M^2}.
\label{eq:neumann}
\end{equation}
At the end of the reaction zone where $y=z=0$, $y_p=y_0$, the square root vanishes and $w$ equals the local sound speed $(1+\gamma M^2)/(\gamma+1)M$: this is the sonic CJ point. Finally, substituting (\ref{eq:Dsteady}) into (\ref{eq:nd3}) and cancelling the common factor $\Pi$ gives the equations for the species,
\begin{equation}
w\frac{\dd y}{\dd s} = -\Omega, \qquad w\frac{\dd z}{\dd s} = \Omega - z, \qquad y(0)=y_0,\quad z(0)=0,
\label{eq:zndode}
\end{equation}
in which $w$ and $T$ (and hence $\Omega$) are known functions of $y_p=y_0-y-z$ through (\ref{eq:wT}). These two ordinary differential equations determine the structure of the wave.

\subsection{Structure}

\begin{figure}[!tb]
\centering
\includegraphics[width=\textwidth]{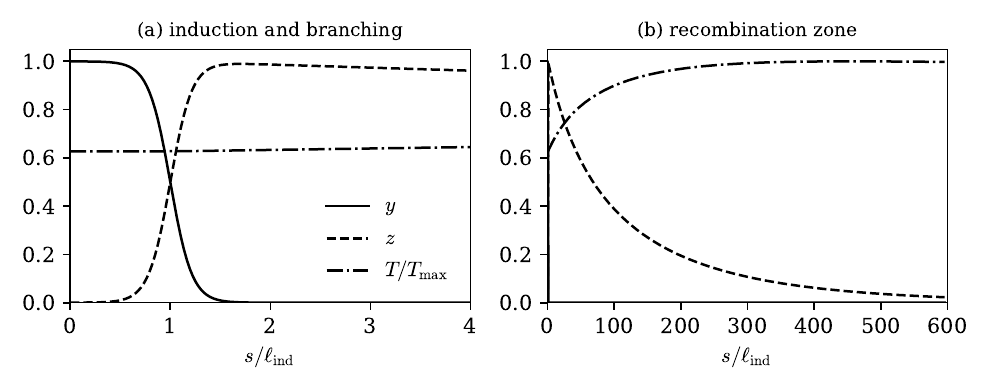}
\caption{Structure of the planar CJ detonation with ZLD kinetics for the parameter values (\ref{eq:values}): fuel $y$, chain carrier $z$ and temperature $T/T_{\max}$ against distance $s$ behind the shock, in units of the induction length. (a) Induction zone and branching layer. (b) Recombination zone.}
\label{fig:znd}
\end{figure}

Figure~\ref{fig:znd} shows the solution of (\ref{eq:zndode}) for $y_0=1$ and the parameter values (\ref{eq:values}). The structure has three parts. Immediately behind the shock there is an \emph{induction zone}, in which the chain carrier grows exponentially from the small level set by initiation. The fuel is barely consumed and the temperature stays at its Neumann value, because no heat is released while $z$ is small. The induction zone ends in a thin \emph{branching layer}, in which the fuel is converted almost entirely into chain carrier, still with little heat release. There follows a long \emph{recombination zone}, in which the chain carrier recombines and the temperature rises from $T_n=5.8$ to the CJ value $\TCJ=9$. We define the induction length $\lind$ as the distance from the shock to the point where half the fuel has been consumed, $y=y_0/2$. For the incoming wave,
\begin{equation}
\hat\ell_1 = \frac{k_C\,\lind}{c_0} = 0.01343 .
\label{eq:l1}
\end{equation}
We define the recombination length $\lrec$ as the distance beyond $\lind$ over which $1-\e^{-1}$ of the heat is released. It is $\lrec\approx1.44\,c_0/k_C\approx107\,\lind$.

\subsection{The induction length and Li\~{n}\'{a}n's regimes}

In the induction zone $z$ is small, $y\approx y_0$ and $T\approx T_n$. The chain-carrier equation in (\ref{eq:zndode}) is then linear, and $z$ grows like $\exp(\sigma_ns/w_n)$, where
\begin{equation}
\sigma_n \equiv y_0\exp\left[\beta\left(\frac1{T_c}-\frac1{T_n}\right)\right] - 1
\label{eq:sigma}
\end{equation}
is the net growth rate of the chain carrier at the Neumann state in units of $k_C$: branching minus recombination. It is positive precisely when $T_n>T_c(y_0)$, the crossover temperature (\ref{eq:Tcy}) of the mixture: the shock must heat the gas above crossover for the chain to branch. For the incoming wave $\sigma_n\approx685$. Appendix~\ref{app:induction} shows that, for $\sigma_n\gg1$,
\begin{equation}
\lind \simeq \frac{w_n}{\sigma_n}\ln\frac{y_0}{\varepsilon}\;\frac{c_0}{k_C},
\label{eq:lindasym}
\end{equation}
which gives $\hat\ell_1\simeq0.01345$, within $0.1$ per cent of (\ref{eq:l1}). The induction zone is thermally neutral, so (\ref{eq:lindasym}) contains no compressibility factor of the kind that appears for one-step kinetics, derived by He \& Clavin~\cite{HeClavin1994}.

The quantity $1+\sigma_n$ is the ratio of branching to recombination rates at the Neumann state. In Li\~{n}\'{a}n's terminology~\cite{Linan1971}, the incoming wave, with $\sigma_n\gg1$, is in the regime of \emph{slow recombination}. The chain carrier accumulates before it recombines, and the heat is released over a zone longer than the induction zone by a factor of order $\sigma_n/\ln(1/\varepsilon)$. As the wave enters leaner mixture, $\sigma_n$ decreases. Recombination becomes comparable with branching inside the induction zone, and at $\sigma_n=0$ branching can no longer sustain the chain.

\subsection{Stability}

Short \& Sharpe~\cite{ShortSharpe2003} showed for a related two-step model that the planar wave is stabilised as the ratio of heat-release to induction lengths increases; similar conclusions hold for three-step chain-branching kinetics~\cite{Ng2005}. With $\lrec/\lind\approx107$ the base wave is expected to be stable, and the computations of \S\ref{sec:unsteady} confirm this. Starting from the steady wave in uniform mixture with the shock speed perturbed by one per cent, the perturbation decays within a few induction times. No growing oscillation appears over $600$ induction lengths of propagation, about six recombination lengths. With uniform fuel the layer thickness plays no role, so this conclusion is independent of $\Lambda$.

\section{The supported wave}\label{sec:support}

\subsection{Characteristics, sonic isolation and regimes}\label{sec:regimes}

\begin{figure}[!tb]
\centering
\includegraphics[width=\textwidth]{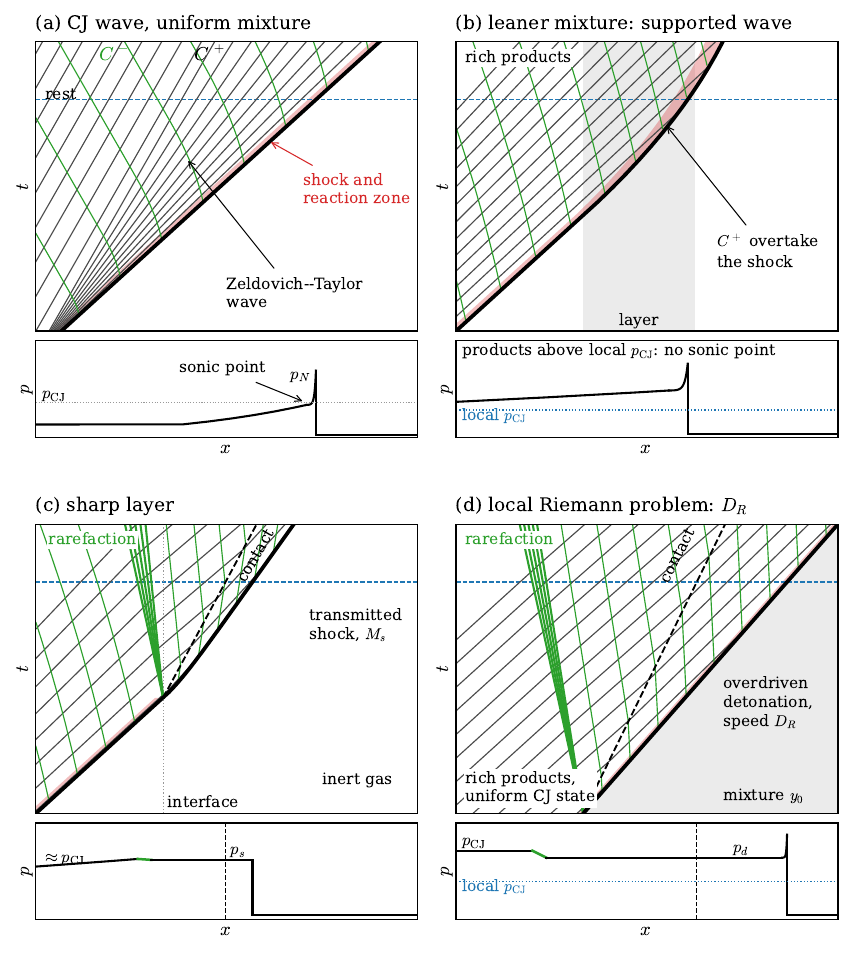}
\caption{Schematic of the wave structure. In each panel, top: the $(x,t)$ plane, with the shock (thick line), the reaction zone (shaded red), $C^+$ characteristics (grey), $C^-$ characteristics and rarefactions (green), and contact surfaces (dashed); bottom: the pressure along the horizontal dashed line. (a) A CJ detonation started from a closed end in uniform mixture: the leading $C^+$ characteristic of the Zeldovich--Taylor wave runs with the shock, which is isolated by its sonic point. (b) The same wave, started far upstream, entering leaner mixture (grey band): the shock slows and is overtaken by $C^+$ characteristics from the rich products, whose pressure stays above the local CJ pressure. (c) A sharp interface: rarefaction, contact surface and transmitted shock. (d) The local Riemann problem of \S\ref{sec:envelope} for $y_0=0.5$: rarefaction, contact surface and overdriven detonation of speed $D_R$, overtaken by $C^+$ characteristics; the pressure behind it, $p_d$, exceeds the local CJ pressure. Slopes and pressure levels are those of the reference case $\MCJ=5$, $\gamma=1.4$; time is scaled so that the incoming CJ shock has unit slope.}
\label{fig:schematic}
\end{figure}

Wherever the gas has finished reacting and its entropy is uniform, the flow is governed by the classical theory of characteristics~\cite{ZeldovichRaizer1966,LandauLifshitz1987,ClavinSearby2016}. The Riemann invariants
\begin{equation}
J_\pm = u \pm \frac{2c}{\gamma-1}
\label{eq:Riemanninv}
\end{equation}
are then constant along the forward and backward characteristics, $\dd x/\dd t=u\pm c$, where $c=(\gamma p/\rho)^{1/2}$ is the local sound speed. We call these the $C^+$ and $C^-$ characteristics. Figure~\ref{fig:schematic} shows four configurations in the $(x,t)$ plane, each with the pressure profile at one instant beneath it.

\emph{(a) CJ wave in uniform mixture.} Behind a CJ detonation started from a closed end, the products form a centred rarefaction, the Zeldovich--Taylor wave~\cite{Zeldovich1940,Taylor1950,ZeldovichRaizer1966}. In it the $C^+$ characteristics are straight lines through the point of initiation, and $J_-$ is uniform (Appendix~\ref{app:ZT}). At the CJ point the gas moves away from the shock at exactly the sound speed, $u+c=D$, so the leading $C^+$ characteristic runs parallel to the shock. No characteristic from the products can catch the shock, which is therefore isolated from everything behind its reaction zone. This sonic isolation is what makes the CJ speed depend only on the mixture ahead.

\emph{(b) Leaner mixture.} When the shock enters leaner mixture, the heat released behind it decreases and the shock slows. The $C^+$ characteristics leaving the rich products still travel at almost the old CJ speed. They now overtake the shock, carrying with them the high pressure of products formed from rich mixture. The pressure behind the reaction zone stays above the local CJ pressure, the gas no longer leaves the reaction zone sonically, and the wave is overdriven relative to the local CJ state. The products support the wave.

\emph{(c) Sharp layer.} If the fuel vanishes abruptly, the shock leaves the reaction zone behind. Once the remaining reaction is complete, the flow near the interface is a Riemann problem: a rarefaction propagating back into the products, a contact surface separating products from inert gas, and a transmitted shock (\S\ref{sec:sharp}).

\emph{(d) Local Riemann problem.} Uniform rich products at their CJ state are placed in contact with quiescent mixture of fuel content $y_0$. The solution is self-similar: a rarefaction into the products, a contact surface, and an overdriven detonation of speed $D_R$. Behind the detonation $u+c>D_R$, so the $C^+$ characteristics from the rich products reach it, as in panel (b). This idealised problem supplies the reference speed of the supported wave (\S\ref{sec:envelope}).

Whether the support can be removed depends on the Zeldovich--Taylor wave. As the shock crosses the layer, the local CJ pressure falls by an amount comparable with itself in the crossing time, of order $a/D$. For the front to follow the local CJ state, the pressure of the products immediately behind it must fall by a comparable amount in the same time. The Zeldovich--Taylor wave is self-similar: its state depends only on $(x-x_w)/t$, where $x_w$ is the closed end and $t$ the time since initiation. At a fixed distance behind the front, its state therefore changes on the time scale $L_0/D$, and in the crossing time it changes by a relative amount of order $a/L_0$. Sonic isolation can be restored only if $L_0\ll a$, that is, if the wave has travelled a distance short compared with the thickness of the layer.

\begin{figure}[!tb]
\centering
\includegraphics[width=0.62\textwidth]{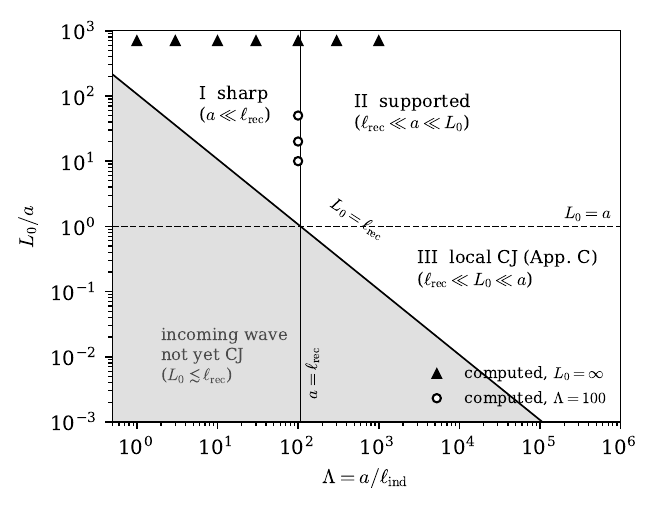}
\caption{Regimes in the plane of $\Lambda=a/\lind$ and $L_0/a$, for $\lrec/\lind\approx107$. In the shaded region the incoming wave has not yet become a CJ wave. Symbols mark the computations of \S\ref{sec:unsteady}: the piston-supported sweep ($L_0=\infty$, triangles) and the runs with a Zeldovich--Taylor wave at $\Lambda=100$, $L_0/a=10$, $20$ and $50$ (circles).}
\label{fig:regimes}
\end{figure}

These considerations involve four lengths: the induction length $\lind$, the recombination length $\lrec\approx107\lind$, the layer thickness $a$, and the length $L_0$ of the Zeldovich--Taylor wave. The incoming wave is a CJ wave only if $\lrec\ll L_0$; otherwise it has not yet settled after initiation. The two remaining comparisons, of $a$ with $\lrec$ and of $a$ with $L_0$, define three regimes, shown in figure~\ref{fig:regimes}.
\begin{enumerate}
\item[I.] $a\ll\lrec$ (\emph{sharp layer}). The shock leaves the reactive mixture before its reaction zone can respond, as in figure~\ref{fig:schematic}(c). It goes over to the transmitted shock of \S\ref{sec:sharp}.
\item[II.] $\lrec\ll a\ll L_0$ (\emph{supported wave}). The reaction zone adjusts to the local mixture, but the products behind it do not relax, as in figure~\ref{fig:schematic}(b). The wave is overdriven, and its speed is referred to the envelope $D_R$ of \S\ref{sec:envelope}.
\item[III.] $\lrec\ll L_0\ll a$ (\emph{local CJ}). Both the reaction zone and the products adjust, and the wave passes through local CJ states until its induction zone decouples. This regime requires the wave to have travelled less than the thickness of the layer. It is outside the main line of this paper and is treated in Appendix~\ref{app:localCJ}.
\end{enumerate}
In terms of $\Lambda$ and $L_0/a$, the boundaries are $\Lambda=\lrec/\lind$, $L_0/a=1$ and $\Lambda L_0/a=\lrec/\lind$. For a detonation crossing a cloud, $L_0$ is comparable with the size of the cloud and much larger than the thickness of its edge, so $L_0/a\gg1$. Regimes I and II are therefore the cases of practical interest, and in both the description in terms of local CJ states does not apply.

\subsection{The supported wave as a local Riemann problem}\label{sec:envelope}

We now ask what speed the supported wave tends to. In regime II two simplifications hold. First, because $a\ll L_0$, the Zeldovich--Taylor wave hardly changes while the front crosses the layer, so the products behind the front may be taken as uniform, at the CJ state of the rich mixture. Second, because $\lrec\ll a$, the fuel content ahead of the shock changes only slightly over the length of the reaction zone. Near any position $x_s$ the front therefore sees an effectively uniform mixture, of fuel content $y_0=\tfrac12\erfc x_s$. Under these two conditions the wave at each position is approximated by the solution of an idealised problem. Uniform rich products, at their CJ state, are in contact with quiescent mixture of uniform fuel content $y_0$. We seek the state that this configuration reaches once all transients have decayed. This is a Riemann problem: its long-time solution is self-similar and consists of uniform states separated by waves. Its leading wave travels at a speed $D_R(y_0)$, which plays for the supported wave the role that the local CJ speed $\MCJ(y_0)$ plays in the local-CJ description of Appendix~\ref{app:localCJ}. The wave in the layer can be expected to approach $D_R(y_0(x_s))$ only after it has travelled several recombination lengths in mixture of nearly constant composition. The comparison with computations in \S\ref{sec:unsteady} will show how far that expectation holds.

The left state of the Riemann problem is the CJ state of the rich products. From (\ref{eq:RH}) and (\ref{eq:wT}) with $y_0=1$ and $y_p=1$, together with $u=\MCJ-w$, it is
\begin{equation}
\pCJ = \frac{1+\gamma\MCJ^2}{\gamma(\gamma+1)}, \qquad
\rho_{\mathrm{CJ}} = \frac{(\gamma+1)\MCJ^2}{1+\gamma\MCJ^2}, \qquad
\uCJ = \frac{\MCJ^2-1}{(\gamma+1)\MCJ}, \qquad \cCJ^2=\frac{\gamma\pCJ}{\rho_{\mathrm{CJ}}},
\label{eq:CJstate}
\end{equation}
where $\MCJ=\MCJ(1)$ is the Mach number of the incoming wave. The right state is quiescent mixture of fuel content $y_0$. The solution, sketched in figure~\ref{fig:schematic}(d), has the same three-wave pattern as the sharp-layer flow of figure~\ref{fig:schematic}(c), except that the leading wave is now a detonation rather than an inert shock.
\begin{enumerate}
\item[(i)] A rarefaction travels back into the rich products. It is a backward-facing simple wave, so the forward invariant $J_+$ is the same throughout it, and the entropy is unchanged. A product state of pressure $p$ reached through the rarefaction therefore has sound speed $c=\cCJ(p/\pCJ)^{(\gamma-1)/2\gamma}$ and, from $J_+=\uCJ+2\cCJ/(\gamma-1)$, velocity
\begin{equation}
u = \uCJ + \frac{2\cCJ}{\gamma-1}\left[1-\left(\frac{p}{\pCJ}\right)^{(\gamma-1)/2\gamma}\right].
\label{eq:rarefaction}
\end{equation}
\item[(ii)] A contact surface separates the rich products from the gas processed by the leading wave. Pressure and velocity are continuous across it.
\item[(iii)] The leading wave is a detonation of speed $D_R$ in the mixture of fuel content $y_0$. Over the long time considered, its reaction zone is thin and the mixture behind it has reacted completely. Its state therefore follows from (\ref{eq:RH}) and (\ref{eq:wgeneral}) with $M=D_R$ and $y_p=y_0$. The gas velocity is $u_d=D_R-w$, and the pressure is $p_d=1/\gamma+D_R^2-D_Rw$.
\end{enumerate}
Equating pressure and velocity on the two sides of the contact surface, that is, setting $p=p_d$ and $u=u_d$ in (\ref{eq:rarefaction}), gives one equation for $D_R$:
\begin{equation}
u_d(D_R; y_0) = \uCJ + \frac{2\cCJ}{\gamma-1}\left[1-\left(\frac{p_d(D_R;y_0)}{\pCJ}\right)^{(\gamma-1)/2\gamma}\right].
\label{eq:DR}
\end{equation}
The square root in (\ref{eq:wgeneral}) is real only if $D_R\ge\MCJ(y_0)$, so the root must satisfy this inequality. The leading detonation is then overdriven relative to the local CJ state, and its state of complete reaction lies on the strong branch of the Hugoniot curve. If (\ref{eq:DR}) had no such root, the products would not support the wave, and the leading wave would be the local CJ detonation, $D_R=\MCJ(y_0)$. For $y_0<1$ the rich products have the higher pressure, and a root always exists.

\begin{figure}[!tb]
\centering
\includegraphics[width=0.55\textwidth]{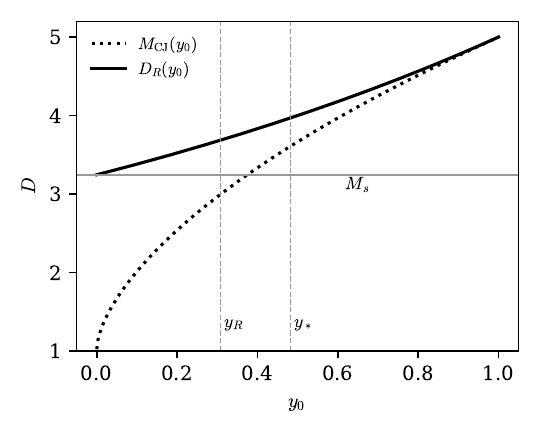}
\caption{The local CJ Mach number $\MCJ(y_0)$ (dotted) and the Riemann envelope $D_R(y_0)$ from (\ref{eq:DR}) (solid), with the detonability limits $y_*$ (local CJ, Appendix~\ref{app:localCJ}) and $y_R$ (supported wave, (\ref{eq:yR})), and the sharp-layer value $M_s$ of \S\ref{sec:sharp}.}
\label{fig:supported}
\end{figure}

We call $D_R(y_0)$ the \emph{Riemann envelope}. Figure~\ref{fig:supported} compares it with $\MCJ(y_0)$. The envelope lies well above the local CJ speed. At $y_0=0.5$, $D_R=4.00$ whereas $\MCJ=3.67$; at $y_0=0.2$, $D_R=3.52$ whereas $\MCJ=2.54$. As $y_0\to0$ the leading wave becomes an inert shock, and $D_R$ tends to the transmitted-shock Mach number $M_s=3.25$ of \S\ref{sec:sharp}.

The supported wave also has a detonability limit. Its shock heats the gas to $T_n(D_R)$, given by (\ref{eq:neumann}) with $M=D_R$, while chain branching requires $T_n>T_c(y_0)$ by (\ref{eq:Tcy}). The two meet at
\begin{equation}
y_R = 0.31, \qquad x_R = 0.35, \qquad D_R(y_R)=3.69 .
\label{eq:yR}
\end{equation}
This lies deeper into the layer than the limit $y_*=0.48$ of a local CJ wave (Appendix~\ref{app:localCJ}), because the support keeps the shock hotter. Beyond $y_R$ the leading wave can no longer ignite the mixture. It is a shock driven by the expanding products, and $M_s$ becomes the appropriate reference.

Two qualifications are needed. First, $D_R$ is the speed of the long-time state of an idealised problem. It is a hydrodynamic reference, not a pointwise bound on every unsteady solution, although all the histories computed in \S\ref{sec:unsteady} approach it from above. Second, (\ref{eq:DR}) assumes complete reaction of the local mixture. Near $y_R$ the induction length becomes long, and complete reaction may not occur on the hydrodynamic time scale, so $y_R$ is a formal limit.

\subsection{The sharp layer}\label{sec:sharp}

When the layer is thin compared with the heat-release zone ($a\ll\lrec$, regime I), the Riemann problem of \S\ref{sec:envelope} is no longer an idealisation applied locally. It arises literally, at the interface itself. The shock passes through the layer before its structure can change, and since the reactant and the inert gas have the same thermodynamic properties, the shock does not notice the interface. What changes is that the gas it compresses from then on contains no fuel. The reaction zone, now detached from the shock, acts on it as a decelerating piston. Once the gas behind has finished reacting, the flow near the interface is the Riemann problem between the rich products and inert gas, sketched in figure~\ref{fig:schematic}(c): a rarefaction into the products, a contact surface, and a transmitted shock. This is (\ref{eq:DR}) with $y_0=0$, and the transmitted shock has Mach number $M_s=D_R(0)$. For an inert shock the state behind is given by the shock relations $u_s=2(M_s-M_s^{-1})/(\gamma+1)$ and $p_s=[1+2\gamma(M_s^2-1)/(\gamma+1)]/\gamma$~\cite{LandauLifshitz1987,Toro2009}. With these, (\ref{eq:DR}) becomes
\begin{equation}
\frac{2}{\gamma+1}\left(M_s-\frac1{M_s}\right) = \uCJ + \frac{2\cCJ}{\gamma-1}\left[1-\left(\frac{p_s}{\pCJ}\right)^{(\gamma-1)/2\gamma}\right].
\label{eq:Riemann}
\end{equation}
For $\MCJ=5$ we find $M_s=3.25$, with post-shock temperature $T_s=2.98$, below $T_c$. The ratio $M_s/\MCJ$ falls from $0.72$ at $Q=1$ to $0.63$ at $Q=50$ (table~\ref{tab:Ms}).

\begin{table}
\caption{The transmitted shock in the sharp-layer limit, for $\gamma=1.4$: CJ Mach number, transmitted-shock Mach number $M_s$ from (\ref{eq:Riemann}), their ratio, and the temperature $T_s$ behind the transmitted shock.}
\centering
\begin{tabular}{rcccc} \toprule
$Q$ & $\MCJ$ & $M_s$ & $M_s/\MCJ$ & $T_s$ \\ \midrule
1    & 2.414  & 1.741 & 0.721 & 1.49 \\
2    & 3.146  & 2.154 & 0.685 & 1.82 \\
5.76 & 5.000  & 3.246 & 0.649 & 2.98 \\
10   & 6.479  & 4.141 & 0.639 & 4.27 \\
20   & 9.055  & 5.719 & 0.632 & 7.30 \\
50   & 14.212 & 8.909 & 0.627 & 16.38 \\ \midrule
$\infty$ & -- & -- & 0.6235 & -- \\ \bottomrule
\end{tabular}
\label{tab:Ms}
\end{table}

The ratio has a simple strong-detonation limit. For $Q\to\infty$, (\ref{eq:MCJ}) gives $\MCJ\simeq2Q^{1/2}\to\infty$. The leading terms of (\ref{eq:CJstate}) are then $\pCJ\simeq\MCJ^2/(\gamma+1)$, $\uCJ\simeq\MCJ/(\gamma+1)$ and $\cCJ\simeq\gamma\MCJ/(\gamma+1)$. The transmitted shock is also strong, with $u_s\simeq2M_s/(\gamma+1)$ and $p_s\simeq2M_s^2/(\gamma+1)$. Writing $M_s=C\MCJ$ and dividing (\ref{eq:Riemann}) by $\MCJ/(\gamma+1)$ gives
\begin{equation}
2C-1 = \frac{2\gamma}{\gamma-1}\left[1-\left(2C^2\right)^{(\gamma-1)/2\gamma}\right],
\label{eq:Cgamma}
\end{equation}
which is independent of the heat release and of the kinetics. Its root is $C=0.6235$ for $\gamma=1.4$, and it hardly depends on $\gamma$. As $\gamma\to1$, (\ref{eq:Cgamma}) becomes $2C-1=-\ln(2C^2)$, with root $C=0.6242$. As $\gamma\to\infty$ it becomes $2C-1=2(1-\sqrt2\,C)$, so that $C=3/(2+2\sqrt2)=0.6213$. For every polytropic gas, therefore, the shock emerging from a sharp layer travels at about $0.62$ of the speed of a strong CJ detonation, and its pressure is $2C^2\approx0.78$ of the CJ pressure.

\subsection{Generality of the support}

The support is a hydrodynamic effect, and it is present at every Mach number. Table~\ref{tab:support} gives $D_R$ and $\MCJ$ at two fuel contents for a range of $Q$. At half the fuel the overdrive $D_R/\MCJ(y_0)-1$ grows from $4$ per cent at $\MCJ\approx2$ to $11$ per cent at $\MCJ\approx14$. At $y_0=0.2$ it exceeds $10$ per cent in every case shown. The Mach number changes the size of the effect but not its nature.

\begin{table}
\caption{The Riemann envelope $D_R$ and the local CJ Mach number $\MCJ(y_0)$ at two fuel contents, for a range of heat-release parameters ($\gamma=1.4$).}
\centering
\begin{tabular}{rccccccc} \toprule
 & & \multicolumn{3}{c}{$y_0=0.5$} & \multicolumn{3}{c}{$y_0=0.2$} \\ \cmidrule(lr){3-5}\cmidrule(lr){6-8}
$Q$ & $\MCJ(1)$ & $\MCJ(y_0)$ & $D_R$ & $D_R/\MCJ(y_0)-1$ & $\MCJ(y_0)$ & $D_R$ & $D_R/\MCJ(y_0)-1$ \\ \midrule
0.5  & 1.932  & 1.618  & 1.682  & 0.039 & 1.365 & 1.555 & 0.139 \\
1    & 2.414  & 1.932  & 2.037  & 0.055 & 1.543 & 1.851 & 0.200 \\
2    & 3.146  & 2.414  & 2.585  & 0.071 & 1.816 & 2.314 & 0.274 \\
5.76 & 5.000  & 3.667  & 4.001  & 0.091 & 2.540 & 3.524 & 0.387 \\
20   & 9.055  & 6.479  & 7.148  & 0.103 & 4.236 & 6.244 & 0.474 \\
50   & 14.212 & 10.099 & 11.179 & 0.107 & 6.479 & 9.741 & 0.503 \\ \bottomrule
\end{tabular}
\label{tab:support}
\end{table}
\section{Unsteady computations}\label{sec:unsteady}

\subsection{Equations in the shock-attached frame}

To follow the shock without smearing it, we solve the equations in coordinates attached to it. Let $x_s(t)$ be the shock position and $D(t)=\dd x_s/\dd t$ its speed, both unknown, and let $\xi=x-x_s(t)$, so that the shock is at $\xi=0$ and the gas behind it occupies $\xi<0$. Since $\partial/\partial t|_x=\partial/\partial t|_\xi-D\,\partial/\partial\xi$, every conservation law $\partial U/\partial t+\partial F/\partial x=S$ becomes $\partial U/\partial t+\partial(F-DU)/\partial\xi=S$. Writing (\ref{eq:nd1})--(\ref{eq:nd3}) in conservation form, we obtain
\begin{equation}
\frac{\partial\boldsymbol U}{\partial t} + \frac{\partial}{\partial\xi}\big[\boldsymbol F(\boldsymbol U) - D\,\boldsymbol U\big] = \boldsymbol S,
\label{eq:conservative}
\end{equation}
with
\begin{equation}
\boldsymbol U = \begin{pmatrix}\rho\\ \rho u\\ \rho E\\ \rho y\\ \rho z\\ \rho f\end{pmatrix},\qquad
\boldsymbol F = \begin{pmatrix}\rho u\\ \rho u^2+p\\ (\rho E+p)u\\ \rho u y\\ \rho u z\\ \rho u f\end{pmatrix},\qquad
\boldsymbol S = \Lambda\hat\ell_1\begin{pmatrix}0\\ 0\\ \dfrac{2Q}{\gamma^2-1}\rho z\\ -\rho\,\Omega\\ \rho(\Omega-z)\\ 0\end{pmatrix}.
\label{eq:UFS}
\end{equation}
Here $E=p/[(\gamma-1)\rho]+u^2/2$ is the total energy per unit mass and $u$ remains the laboratory velocity. The energy equation in (\ref{eq:UFS}) is equivalent to (\ref{eq:nd2}), and the factor $\Lambda\hat\ell_1$ is $\Pi$ by (\ref{eq:Lambda}). The extra variable $f$ is a passive scalar recording the fuel content each particle had when it was shocked. Because particles shocked at different positions start with different amounts of fuel, the burnt fraction $1-y/f$, rather than $y$ itself, is used to measure the induction distance behind the shock.

\subsection{The shock-change equation}

Immediately behind the shock, at $\xi=0^-$, the Rankine--Hugoniot conditions for a shock of Mach number $D$ moving into gas at rest give
\begin{equation}
\rho_s = \frac{(\gamma+1)D^2}{(\gamma-1)D^2+2}, \quad
u_s = \mathcal U(D) = \frac{2}{\gamma+1}\left(D-\frac1D\right), \quad
p_s = \mathcal P(D) = \frac{2\gamma D^2-(\gamma-1)}{\gamma(\gamma+1)},
\label{eq:RHshock}
\end{equation}
together with $y_s=f_s=\tfrac12\erfc x_s$ and $z_s=0$. Steps 0 and I are thermally neutral, so these conditions do not depend on the fuel content. One more relation is needed to determine $D(t)$. It is supplied by the $C^+$ characteristic, which carries information from the reaction zone to the shock. Multiplying the momentum equation in (\ref{eq:nd1}) by $\rho c$ and adding it to the pressure form of the energy equation, $Dp/Dt+\gamma p\,\partial u/\partial x=(\gamma-1)\rho\,\dot q$, where $\dot q=2Q\Pi z/(\gamma^2-1)$ is the rate of heat release per unit mass, gives
\begin{equation}
\left(\frac{\partial}{\partial t}+(u+c)\frac{\partial}{\partial x}\right)p + \rho c\left(\frac{\partial}{\partial t}+(u+c)\frac{\partial}{\partial x}\right)u = (\gamma-1)\rho\,\dot q .
\label{eq:Cplus}
\end{equation}
At the shock, $p_s$ and $u_s$ are functions of $D$ alone, so their rates of change along the shock path, $\dd/\dd t=\partial_t+D\partial_x$, are $\mathcal P'(D)\,\dd D/\dd t$ and $\mathcal U'(D)\,\dd D/\dd t$. The derivative along the characteristic in (\ref{eq:Cplus}) differs from the derivative along the shock path by $(u+c-D)\,\partial/\partial x$. Substituting, and using $\dot q=0$ at the shock where $z_s=0$, gives the shock-change equation
\begin{equation}
\big[\mathcal P'(D) + \rho_sc_s\,\mathcal U'(D)\big]\frac{\dd D}{\dd t} = -\big(u_s+c_s-D\big)\left(\frac{\partial p}{\partial\xi}+\rho_sc_s\frac{\partial u}{\partial\xi}\right)_{\xi=0^-},
\label{eq:shockchange}
\end{equation}
with $\mathcal P'(D)=4D/(\gamma+1)$ and $\mathcal U'(D)=2(1+D^{-2})/(\gamma+1)$~\cite{Williams1985,KasimovStewart2004,Henrick2006}. The flow behind the shock is subsonic relative to it, so $u_s+c_s-D>0$, and (\ref{eq:shockchange}) expresses the acceleration of the shock in terms of the gradients that arrive from behind.

\subsection{Initial and rear boundary conditions}

The domain is $-L\le\xi\le0$, with $L$ several recombination lengths. At $t=0$ the solution is the steady CJ structure of \S\ref{sec:CJ} combined with the Zeldovich--Taylor wave. Because $\lrec\ll L_0$, we use the composite of the two: the ZND profiles of $u$ and $c$ plus their Zeldovich--Taylor values minus the common CJ values, with the entropy and mass fractions taken from the ZND structure (Appendix~\ref{app:ZT}).

At $\xi=-L$, quantities carried out of the domain, namely the backward invariant $J_-$, the entropy and the mass fractions, are extrapolated from the interior. When the $C^+$ characteristic enters the domain ($u-D+c>0$), the forward invariant $J_+$ is taken from the Zeldovich--Taylor wave. In the limit $L_0\to\infty$ the products behind the CJ point are uniform at the CJ state, and all quantities are extrapolated. We call this limit the \emph{piston-supported} case, since the products then act as an inexhaustible reservoir.

\subsection{Numerical method}

The interior equations are discretised with fifth-order weighted essentially non-oscillatory reconstruction and global Lax--Friedrichs flux splitting~\cite{JiangShu1996}, and integrated in time with the third-order strong-stability-preserving Runge--Kutta scheme~\cite{ShuOsher1988}. The shock-change equation is advanced simultaneously, with the gradients at $\xi=0^-$ evaluated by one-sided fourth-order differences. Ghost values ahead of the shock node are obtained by quadratic extrapolation. The time step is limited by the CFL condition and by the branching rate. With twenty grid points per induction length the steady CJ wave is preserved to $4\times10^{-5}$ in $D$. At $\Lambda=10$ the shock Mach number at $x_s=5$ agrees to $5\times10^{-5}$ for $10$, $15$ and $25$ points per induction length. The number of grid points is independent of $\Lambda$, while the number of time steps grows in proportion to $\Lambda$.

\subsection{Piston-supported histories}

\begin{figure}[!tb]
\centering
\includegraphics[width=\textwidth]{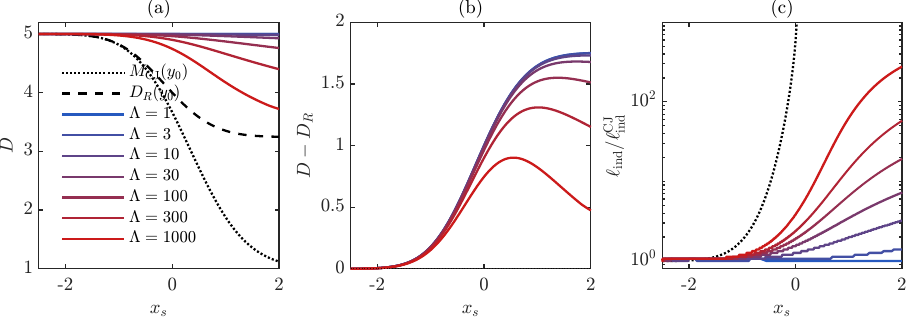}
\caption{Piston-supported computations ($L_0=\infty$) for $1\le\Lambda\le10^3$. (a) Shock Mach number $D(x_s)$, with the local CJ speed $\MCJ(y_0(x_s))$ (dotted) and the Riemann envelope $D_R(y_0(x_s))$ (dashed). (b) The excess $D-D_R$ over the envelope. (c) Induction length relative to its CJ value, with the value $\hat\ell(y_0(x_s))$ of a local CJ wave (dotted).}
\label{fig:sweep}
\end{figure}

Figure~\ref{fig:sweep} shows computed shock histories in the piston-supported case for $1\le\Lambda\le10^3$, together with the local CJ speed and the Riemann envelope of \S\ref{sec:envelope}, and table~\ref{tab:sweep} summarises them.
\begin{enumerate}
\item[(i)] The shock remains far above the local CJ speed throughout. At $x_s=0$, where $\MCJ=3.67$ and $D_R=4.00$, the computed Mach number lies between $5.00$ ($\Lambda\le10$) and $4.75$ ($\Lambda=10^3$). The wave approaches the envelope $D_R$ from above, more closely as $\Lambda$ increases, but it does not reach it within the layer (\S\ref{sec:comparison}).
\item[(ii)] The induction zone does not decouple. A particle shocked at $x_s$ completes its induction after the shock has moved about $\lind$. If $\lind$ grew by an amount comparable with itself over that distance, that is if $\dd\lind/\dd x_s$ reached a value of order unity, the heat release could not keep up with the shock (Appendix~\ref{app:localCJ}). The largest value of $\dd\lind/\dd x_s$ found in the computations grows only from $0.10$ to $0.25$ as $\Lambda$ increases from $3$ to $10^3$. The induction length grows smoothly (figure~\ref{fig:sweep}c), to about $270$ times its upstream value by $x_s=2$ for $\Lambda=10^3$, far below the value $\hat\ell(y_0(x_s))$ of a local CJ wave.
\item[(iii)] Where the position of steepest deceleration lies inside the computed range, it moves upstream as $\Lambda$ increases: $x_s=1.46$, $1.03$ and $0.70$ for $\Lambda=10^2$, $3\times10^2$ and $10^3$.
\end{enumerate}
These computations also test the role of the induction kinetics. Had they controlled the failure of the wave, the decoupling criterion (\ref{eq:criterion}) of Appendix~\ref{app:localCJ} would place it at $x_q=-0.74$, $-0.38$ and $-0.13$ for $\Lambda=10$, $10^2$ and $10^3$. Every computed front passes these positions with its induction zone coupled, and at $x_s=0$ it still travels at $D=4.75$--$5.00$. The front decelerates instead where the support from the burnt gas falls. Its steepest deceleration approaches the centre of the layer, where $D_R$ falls fastest, roughly as $\Lambda^{-1/3}$, and beyond the layer the shock tends to the transmitted-shock Mach number $M_s$, which follows from gas dynamics alone. The decay of a supported wave is therefore governed by its heat-release zone and by the products behind it, not by its induction zone. The kinetics fix the length of the heat-release zone and the limit $y_R$ below which the shock can no longer ignite the mixture, but not where the wave begins to fail.

When the layer is thin, the approach to the transmitted shock is slow. For $\Lambda=10$ the shock Mach number had fallen only to $4.56$ when the shock had travelled $40a$, about four recombination lengths, beyond the layer.

\begin{table}
\caption{Summary of the piston-supported computations: shock Mach number at $x_s=0$ (where $\MCJ=3.67$ and $D_R=4.00$), and the largest value of $\dd\lind/\dd x_s$ and its position.}
\centering
\begin{tabular}{rccc} \toprule
$\Lambda$ & $D(0)$ & $\max\,\dd\lind/\dd x_s$ & at $x_s$ \\ \midrule
3    & 5.000 & 0.10 & 1.66 \\
10   & 4.998 & 0.15 & 1.88 \\
30   & 4.990 & 0.16 & 2.00 \\
100  & 4.963 & 0.17 & 1.99 \\
300  & 4.900 & 0.20 & 1.87 \\
1000 & 4.746 & 0.25 & 1.72 \\ \bottomrule
\end{tabular}
\label{tab:sweep}
\end{table}

\subsection{Comparison with the theory}\label{sec:comparison}

\emph{Supported regime.} Figure~\ref{fig:sweep}(a) compares the piston-supported histories with the Riemann envelope $D_R(y_0(x_s))$ of \S\ref{sec:envelope}, and figure~\ref{fig:sweep}(b) shows the difference $D-D_R$. Every history lies above the envelope. At $x_s=0$, where $D_R=4.00$ and $\MCJ=3.67$, the computed Mach number is $4.96$ for $\Lambda=10^2$, $4.90$ for $\Lambda=3\times10^2$ and $4.75$ for $\Lambda=10^3$. For $\Lambda\ge10^2$ the excess $D-D_R$ first grows, because the envelope falls with the fuel content faster than the shock can decelerate. It then passes through a maximum and decreases. For $\Lambda=10^3$ the maximum is about $0.9$, near $x_s\approx0.5$, and by $x_s=2$ the excess has fallen to $0.47$: there $D=3.72$, against $D_R=3.25$ and $\MCJ=1.12$. The wave thus moves towards the envelope, as \S\ref{sec:envelope} anticipates, but it does not reach it within the layer. The envelope is approached only after the shock has travelled many recombination lengths in mixture of nearly constant composition, whereas even for $\Lambda=10^3$ the layer is only about ten recombination lengths thick. The sharp-layer computation below shows that this relaxation takes about fourteen recombination lengths.

\begin{figure}[!tb]
\centering
\includegraphics[width=0.55\textwidth]{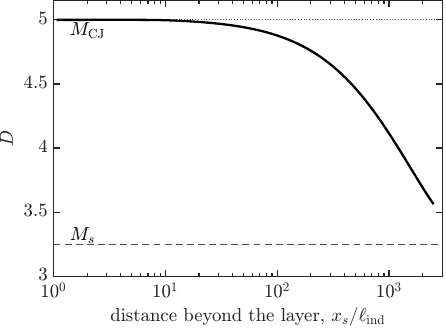}
\caption{Sharp-layer computation ($\Lambda=1$, piston-supported): shock Mach number against distance beyond the layer, in induction lengths, with the CJ value (dotted) and the transmitted-shock Mach number $M_s$ of (\ref{eq:Riemann}) (dashed).}
\label{fig:sharp}
\end{figure}

\emph{Sharp layer.} For $\Lambda=1$ the layer is about a hundred times thinner than the heat-release zone, which places the computation in regime I. Figure~\ref{fig:sharp} follows the shock for $2500$ induction lengths, about $23$ recombination lengths, beyond the layer. The Mach number stays close to $\MCJ=5$ for the first ten induction lengths, while the reaction zone left behind continues to drive the shock. It then decays monotonically: $D=4.65$ at $300$, $4.11$ at $1000$ and $3.56$ at $2500$ induction lengths. Beyond about $300$ induction lengths the decay is exponential,
\begin{equation}
D - M_s \simeq 1.70\,\exp\left(-\frac{x_s}{1490\,\lind}\right),
\label{eq:sharpdecay}
\end{equation}
with $M_s=3.246$ from (\ref{eq:Riemann}). This fit agrees with the computed Mach number to within $0.002$ throughout that range. To test the prediction without assuming it, we fitted $D=M_\infty+A\,\e^{-x_s/L}$ with $M_\infty$, $A$ and $L$ all free. Depending on where the fit begins (at $300$, $600$ or $1000$ induction lengths), it gives $M_\infty=3.26$, $3.24$ and $3.22$, bracketing $M_s=3.246$. The transmitted shock of the Riemann problem is therefore the asymptotic state of the sharp-layer flow. The relaxation length, about $1500\lind\approx14\lrec$, is long, which explains why the shock in the $\Lambda=10$ computation had fallen only to $4.56$ after $40a$, or $400$ induction lengths.

\begin{figure}[!tb][!tb]
\centering
\includegraphics[width=0.55\textwidth]{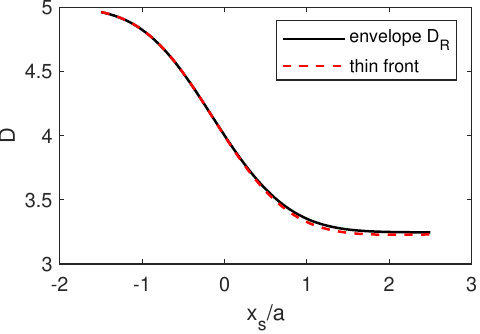}
\caption{Piston-supported wave with the reaction zone collapsed into the front ($\lrec/a\to0$): shock Mach number $D(x_s)$ from the thin-front Euler computation (dashed), and the Riemann envelope $D_R(y_0(x_s))$ of \S\ref{sec:envelope} (solid).}
\label{fig:comparison}
\end{figure}

\emph{Thin-front computations.} 
To separate the gas dynamics from the reaction zone, we also solved the reactive Euler equations with the reaction zone collapsed into the front, so that $\lrec/a\to0$. Behind the front the flow carries the products of complete burning of the local mixture. The shock-attached computation uses a second-order MUSCL--HLL scheme with Heun time stepping. At each step the front speed is found from the compatibility relation $\dd p+\rho c\,\dd u=0$ on the $C^+$ characteristic arriving from behind. When that relation has no root, the front is taken to be sonic. The resulting speeds are shown in figure~\ref{fig:comparison}. For $x_s\le0$ they agree with the envelope $D_R$ to within $0.004$. Beyond that point they lie slightly below it, with $D = 3.997$, $3.568$, $3.327$ and $3.228$ at $x_s=0$, $0.5$, $1$ and $2$, against $D_R = 4.001$, $3.582$, $3.352$ and $3.249$. The largest difference, $0.026$ (about $0.8\,\%$), occurs near $x_s\approx1$--$1.5$. With a thin front, the envelope therefore gives the speed to within $1\,\%$ across the whole layer. The larger departure in the full computations, about $14\,\%$ at $x_s=2$ for $\Lambda=10^3$, is comparable with $\lrec/a\approx0.1$. It is therefore an effect of the finite reaction zone, not a failure of the envelope.

\subsection{The loss of the sonic point}

The mechanism of \S\ref{sec:support} can be checked directly. In the shock-attached frame a $C^+$ characteristic moves relative to the shock with speed $u+c-D$. A steady CJ wave has $u+c-D>0$ in its reaction zone, and $u+c-D\to0$ at its sonic point. The approach is slow, because $u+c-D$ decreases like $(1-y_p/y_0)^{1/2}$ by (\ref{eq:wT}), while the product fraction $y_p$ itself relaxes exponentially over $\lrec$. At $617.5$ induction lengths behind the shock, about six recombination lengths, the steady CJ wave still has $u+c-D=0.322$. The relevant quantity is therefore the excess over this steady value. Table~\ref{tab:sonic} gives it for the piston-supported computation with $\Lambda=100$, at several positions of the shock.

\begin{table}
\caption{Speed $u+c-D$ of the $C^+$ characteristics relative to the shock, $617.5$ induction lengths behind it, for the piston-supported computation with $\Lambda=100$. The excess over the steady CJ value, $0.322$, is compared with $\MCJ-D$.}
\centering
\begin{tabular}{rccccc} \toprule
$x_s$ & $D$ & $u+c-D$ & excess over steady CJ & $\MCJ-D$ \\ \midrule
$-1.0$ & 4.996 & 0.325 & 0.003 & 0.004 \\
$-0.5$ & 4.987 & 0.335 & 0.013 & 0.014 \\
$0.0$  & 4.963 & 0.358 & 0.036 & 0.037 \\
$0.5$  & 4.923 & 0.397 & 0.075 & 0.077 \\
$1.0$  & 4.872 & 0.448 & 0.126 & 0.128 \\
$1.5$  & 4.816 & 0.502 & 0.180 & 0.184 \\ \bottomrule
\end{tabular}
\label{tab:sonic}
\end{table}

The excess agrees with $\MCJ-D$ to within $0.004$ at every position. The products formed from the rich mixture still carry their $C^+$ characteristics at the speed of the incoming CJ wave, $u+c\simeq\MCJ$, while the shock has slowed to $D$. These characteristics therefore outrun the shock by $\MCJ-D$, exactly as figure~\ref{fig:schematic}(b) depicts. In the computation with $\Lambda=10^3$, where the shock decelerates more strongly, $u+c-D$ exceeds $0.68$ everywhere behind the shock for $x_s\ge0$. For $x_s=1$ and $1.5$ its minimum lies inside the domain, $62$ and $144$ induction lengths behind the shock, with values $1.11$ and $1.01$. There is no sonic point, and the rich products communicate with the shock throughout. The situation is the reverse of that analysed by Clavin \emph{et al.}~\cite{Clavin2021} for spherical detonations, in which a detonation survives only if the sonic point in the rarefaction behind it can catch up with the reaction zone.

\subsection{The role of the Zeldovich--Taylor wave}

\begin{figure}[!tb]
\centering
\includegraphics[width=0.8\textwidth]{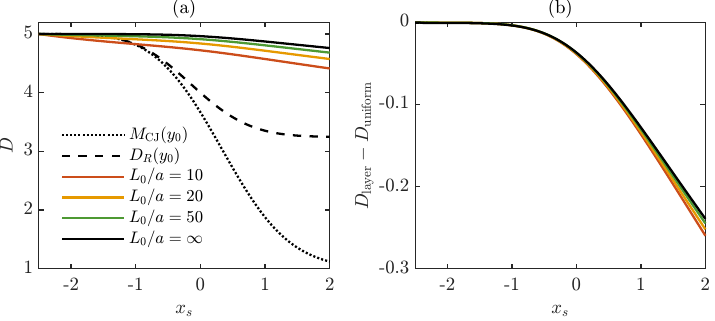}
\caption{Computations at $\Lambda=100$ with a Zeldovich--Taylor wave. (a) Shock Mach number $D(x_s)$ for $L_0/a=10$, $20$, $50$ and $\infty$, with the local CJ speed (dotted) and the Riemann envelope (dashed). (b) The effect of the layer alone, $D_{\rm layer}-D_{\rm uniform}$, obtained by subtracting computations with uniform fuel and the same $L_0$; for $L_0=\infty$, $D_{\rm uniform}=\MCJ$.}
\label{fig:taylor}
\end{figure}

Figure~\ref{fig:taylor}(a) shows histories at $\Lambda=100$ for $L_0/a=10$, $20$, $50$ and $\infty$. In every case $\lrec/L_0\le0.1$. The shock Mach numbers at $x_s=2$ are $4.41$, $4.57$, $4.68$ and $4.76$. At first sight, then, a finite $L_0$ accelerates the decay by an amount of about $3.5\,a/L_0$. Most of this difference, however, has nothing to do with the layer. With $L_0/a=10$ the shock has already slowed to $4.94$ at $x_s=-2$, where $y_0\approx1$. The heat-release zone of the incoming wave lies in the Zeldovich--Taylor gradient, whose length scale is $L_0$, and the wave falls short of the CJ speed by a relative amount of order $\lrec/L_0$.

To separate the two effects we repeated each computation with uniform fuel and the same $L_0$, and took the difference $D_{\rm layer}-D_{\rm uniform}$ (figure~\ref{fig:taylor}b). At $x_s=2$,
\begin{equation}
D_{\rm layer}-D_{\rm uniform} = -0.260,\ -0.252,\ -0.245,\ -0.239 \qquad (L_0/a=10,\ 20,\ 50,\ \infty).
\label{eq:control}
\end{equation}
The decay caused by the layer is thus independent of $L_0$ to leading order. Relative to the piston-supported value, the corrections $0.021$, $0.013$ and $0.006$ are of order $a/L_0$, with a coefficient of about $0.25$. This is what the argument of \S\ref{sec:support} requires. When $a\ll L_0$ the Zeldovich--Taylor wave relaxes the products by only a small fraction during the crossing, so the wave behaves as if it were piston-supported. The piston-supported histories of figure~\ref{fig:sweep} therefore describe the supported regime at leading order, whatever the value of $L_0$.

\section{Concluding remarks}\label{sec:conclusions}

We have considered the last stage in the life of a planar CJ detonation, its passage through a mixing layer in which the fuel runs out. With Zeldovich--Li\~{n}\'{a}n--Dold kinetics the reaction cut-off is physical: it is the crossover temperature, which rises as the fuel is depleted. The incoming wave lies in Li\~{n}\'{a}n's regime of slow recombination, with a heat-release zone about a hundred induction lengths long, and it is stable.

The main conclusion is hydrodynamic. A CJ detonation is isolated from the flow behind it by its sonic point, but this isolation is lost when the chemical energy decreases in the direction of propagation. The products of the rich mixture are then at a higher pressure than the local CJ state, the wave becomes overdriven, and forward characteristics carry the state of the rich products to the shock. Only the Zeldovich--Taylor wave can remove this support, and it does so only if the distance $L_0$ travelled since initiation is small compared with the layer thickness $a$. In the usual case, $a\ll L_0$, the wave does not follow the local CJ states and its induction zone does not decouple. Instead its speed approaches from above the envelope $D_R(y_0)$ given by a Riemann problem between the rich products and the local mixture. The envelope lies $4$ to $50$ per cent above the local CJ speed, depending on the Mach number and the fuel content. It has its own detonability limit $y_R$, and it ends at the transmitted-shock Mach number $M_s$. For strong detonations $M_s/\MCJ$ tends to a constant $C(\gamma)\approx0.62$, almost the same for every polytropic gas. Shock-fitted computations with $\Lambda$ up to $10^3$ confirm that the induction zone does not decouple. They also show directly that no sonic point forms: forward characteristics behind the shock outrun it by $\MCJ-D$. Uniform-fuel controls confirm that the decay caused by the layer is independent of $L_0$ to leading order.

None of these conclusions depends on the kinetic model. The loss of sonic isolation, the Riemann envelope and the limit $C(\gamma)$ follow from gas dynamics alone. The kinetics determine only the length of the reaction zone, which separates the sharp and supported regimes, and the detonability limit $y_R$. The same picture should therefore hold for one-step or detailed chemistry.

Several questions remain. Computations with a thin reactive front, which isolate the gas dynamics from the reaction zone, follow the envelope to within one per cent throughout the layer, lying slightly below it. An evolution equation for a thin supported shock, derived from the characteristic structure of \S\ref{sec:support} in the spirit of the integral equation obtained by Clavin \& Denet~\cite{ClavinDenet2018} for the decay of a detonation after its piston is stopped, would therefore add little to the envelope. The excess of the computed speed over the envelope is instead a reaction-zone effect. For $\Lambda=10^3$ it is about $14$ per cent at $x_s=2a$, comparable with $\ell_{\rm rec}/a\approx0.1$. A theory of it would require an expansion in this ratio. The formulation of weakly exothermic oblique detonations by Mart\'{i}nez-Ruiz \emph{et al.}~\cite{MartinezRuiz2020}, in which the reaction region is described by the linearized Euler equations in characteristic form, suggests one route. The boundary between the sharp and supported regimes might be sharpened by using the local, rather than the upstream, relaxation length of the reaction zone. In diverging geometries the products expand on the scale of the radius $R$ of the front, so $R/a$ should take the place of $L_0/a$. That extension, and the interaction of gradient-induced failure with intrinsic instability when the base wave is unstable, lie beyond the present one-dimensional study.

\appendix
\titleformat{\section}{\centering\normalsize\bfseries}{\MakeUppercase{\appendixname}\ \thesection.}{0.5em}{\MakeUppercase}
\renewcommand{\thesubsection}{\thesection.\arabic{subsection}}

\section{Structure of the induction zone}\label{app:induction}

We derive (\ref{eq:lindasym}) from (\ref{eq:zndode}) for $\sigma_n\gg1$. In the induction zone $z\ll y_0$, so the fuel is undepleted, little heat is released, and $T=T_n$, $w=w_n$. Write $K_n=\exp[\beta(1/T_c-1/T_n)]$ for the branching rate at the Neumann state, so that $\sigma_n=y_0K_n-1$. The equation for $z$ in (\ref{eq:zndode}), with $\Omega=y_0K_n(z+\varepsilon)$, becomes linear,
\begin{equation}
w_n\frac{\dd z}{\dd s} = \sigma_n z + \varepsilon y_0K_n, \qquad\text{whence}\qquad
z = \frac{\varepsilon y_0K_n}{\sigma_n}\left(\e^{\sigma_ns/w_n}-1\right).
\label{eq:zind}
\end{equation}
The chain carrier therefore grows exponentially, on the length scale $w_n/\sigma_n$, from the level $\varepsilon$ set by initiation. Once $z$ becomes comparable with $y_0$, the fuel is consumed in a branching layer whose thickness is also of order $w_n/\sigma_n$. There recombination is slower than branching by a factor $\sigma_n^{-1}$, so $y+z=y_0$ to leading order and the equation for $z$ takes the logistic form
\begin{equation}
w_n\frac{\dd z}{\dd s} = K_n\,(y_0-z)(z+\varepsilon), \qquad
z+\varepsilon = \frac{y_0+\varepsilon}{1+(y_0/\varepsilon)\exp[-K_n(y_0+\varepsilon)s/w_n]},
\end{equation}
the solution satisfying $z(0)=0$. Half the fuel has been consumed, $z=y_0/2$, when
\begin{equation}
s = \frac{w_n}{K_n(y_0+\varepsilon)}\ln\left(\frac{y_0}{\varepsilon}+1\right) \simeq \frac{w_n}{\sigma_n}\ln\frac{y_0}{\varepsilon},
\end{equation}
correct to relative order $\sigma_n^{-1}$ and $\varepsilon$. In dimensional form this is (\ref{eq:lindasym}). Writing $\sigma_n$ in place of $y_0K_n$ accounts for recombination during the exponential phase, and keeps the estimate useful down to $\sigma_n$ of order a few.

\section{The Zeldovich--Taylor wave}\label{app:ZT}

Consider a planar CJ detonation initiated at a closed end $x=x_w$ at time $t=-t_0$. Behind the CJ point the products are isentropic and form a centred simple wave~\cite{Zeldovich1940,Taylor1950,LandauLifshitz1987,ClavinSearby2016}. In a simple wave one Riemann invariant is uniform. Here it is $J_-$, which takes its value at the CJ point:
\begin{equation}
J_- = u - \frac{2c}{\gamma-1} = \uCJ - \frac{2\cCJ}{\gamma-1}.
\end{equation}
The $C^+$ characteristics are straight lines through the point of initiation,
\begin{equation}
u + c = \frac{x-x_w}{t+t_0}.
\end{equation}
Solving these two relations for $u$ and $c$ gives
\begin{equation}
c = \frac{\gamma-1}{\gamma+1}\left(\frac{x-x_w}{t+t_0} - J_-\right), \qquad u = \frac{x-x_w}{t+t_0} - c,
\end{equation}
valid where $u\ge0$. Behind the point at which $u$ vanishes the gas is at rest, with $c=c_w=-(\gamma-1)J_-/2$. The shock at $x_0$ at $t=0$ is a distance $L_0=x_0-x_w$ from the wall, so $t_0=L_0/\MCJ$. The composite initial condition of \S\ref{sec:unsteady} replaces the CJ values of $u$ and $c$ in the ZND structure by their Zeldovich--Taylor values, adding the difference to the ZND profiles. The forward invariant $J_+=u+2c/(\gamma-1)$ of this solution supplies the rear boundary condition.

\section{The local-CJ description}\label{app:localCJ}

If $\lrec\ll L_0\ll a$ (regime III), the products relax as fast as the front advances, and the wave may be expected to propagate as a sequence of steady CJ waves. Each would be appropriate to the fuel content at the current position of the shock,
\begin{equation}
D \simeq \MCJ(y_0), \qquad y_0 = \tfrac12\erfc x_s .
\label{eq:qs}
\end{equation}
This regime requires the wave to have travelled less than the thickness of the layer, which is rarely the case in practice. The local CJ states are nevertheless the natural reference against which the supported wave of \S\ref{sec:support} is measured, and we record their properties here.

\subsection{A detonability limit}

\begin{figure}[!tb]
\centering
\includegraphics[width=\textwidth]{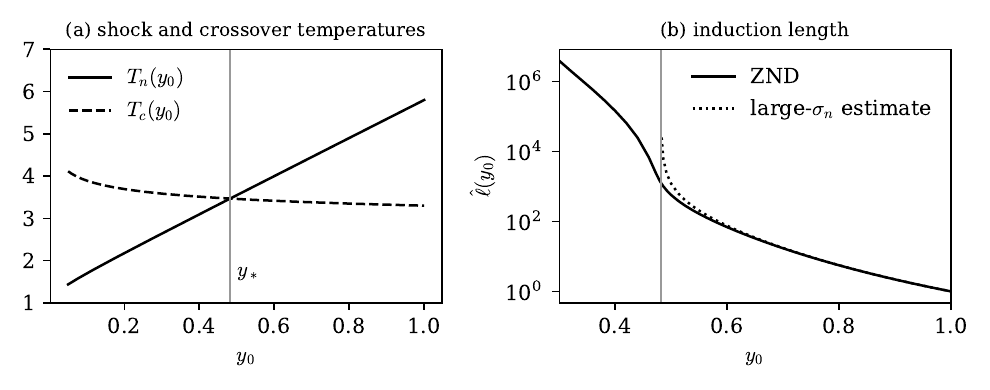}
\caption{Steady CJ waves in a mixture with fuel content $y_0$. (a) Neumann temperature $T_n$ and crossover temperature $T_c$, (\ref{eq:Tcy}), which meet at the detonability limit $y_*=0.482$. (b) The induction length $\hat\ell=\lind(y_0)/\lind(1)$ from (\ref{eq:zndode}) (solid), and the estimate (\ref{eq:lindasym}) (dotted).}
\label{fig:local}
\end{figure}

As $y_0$ decreases, the Neumann temperature $T_n(y_0)$ of the local CJ wave falls and the crossover temperature $T_c(y_0)$ of (\ref{eq:Tcy}) rises (figure~\ref{fig:local}a). They meet where $\sigma_n=0$:
\begin{equation}
T_n\big(\MCJ(y_*)\big) = T_c(y_*).
\label{eq:ystar}
\end{equation}
For the values (\ref{eq:values}), $y_*=0.482$, at $x_*=0.032$, where $\MCJ=3.61$ and $T_n=3.47$. For $y_0<y_*$ a CJ wave cannot be supported by chain branching. This is the analogue for detonations of the flammability limit of flames with linear recombination~\cite{Joulin1985,LeeKim2024}, and of the detonability limit found by Short \& Quirk~\cite{ShortQuirk1997} for uniform mixtures. It is a limit for a chain-branching CJ structure, not an absolute bar to reaction. The steady induction length (figure~\ref{fig:local}b) grows by three orders of magnitude between $y_0=1$ and $y_*$, but it remains finite at $y_*$, because heat released by recombination during the long induction period raises the temperature above crossover. Below $y_*$ it is set by a slow thermal runaway driven by initiation and recombination.

\subsection{Decoupling of the induction zone}

\begin{figure}[!tb]
\centering
\includegraphics[width=0.55\textwidth]{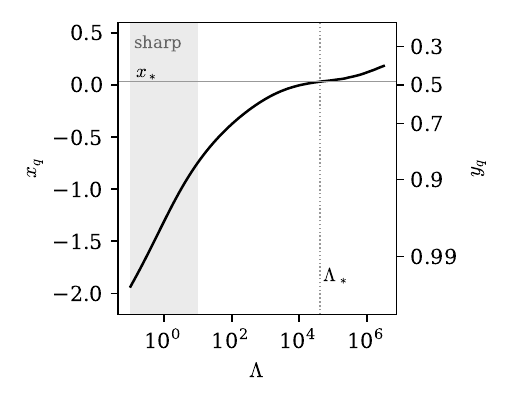}
\caption{Local-CJ prediction of the quenching position $x_q$ and local fuel content $y_q$ (right-hand scale) from (\ref{eq:criterion}). The horizontal line marks the detonability limit $x_*$. The prediction applies only when $\lrec\ll L_0\ll a$; the shaded range is outside it.}
\label{fig:quench}
\end{figure}

Long before $y_*$ is reached, the induction zone of a local CJ wave ceases to be quasi-steady, for a reason analogous to that which governs the decoupling of the induction zone from a decaying shock in direct initiation~\cite{HeClavin1994}. A particle shocked at $x_s$ completes its induction after the shock has moved a distance comparable with $\lind$. If, over that distance, the induction length increases by an amount comparable with itself, the heat release cannot catch up with the shock. The induction zone therefore remains coupled only while $\dd\lind/\dd x_s\ll1$. Taking the quenching position $x_q$ to be where this derivative reaches unity, and writing $\lind=\hat\ell(y_0)\lind(1)$ with $y_0=\tfrac12\erfc x_s$, we obtain
\begin{equation}
\frac{\dd\hat\ell}{\dd x}\bigg|_{x=x_q} = \Lambda, \qquad\text{that is}\qquad
-\frac{\dd\hat\ell}{\dd y_0}\,\frac{\e^{-x_q^2}}{\sqrt\pi} = \Lambda,
\label{eq:criterion}
\end{equation}
where $\e^{-x^2}/\sqrt\pi$ is $-\dd y_0/\dd x$. The left-hand side increases monotonically with $x_q$, so there is one root for each $\Lambda$ (figure~\ref{fig:quench}). For $\Lambda=10^3$ and $10^4$ we find $x_q\approx-0.13$ and $-0.01$, with local fuel contents $0.58$ and $0.50$. The curve crosses $x_*$ at $\Lambda\approx4.1\times10^4$. The criterion is a prediction: it has not been tested by computation, and loss of quasi-steadiness need not imply irreversible failure. In the supported regime, by contrast, the computed values of $\dd\lind/\dd x_s$ never exceed $0.25$ (\S\ref{sec:unsteady}).

\small
\bibliographystyle{unsrt}
\bibliography{final_detonation}

@article{Zeldovich1940,
  author  = {Zel'dovich, Ya. B.},
  title   = {On the theory of the propagation of detonation in gaseous systems},
  journal = {Zh. Eksp. Teor. Fiz.},
  volume  = {10},
  pages   = {542--568},
  year    = {1940},
  note    = {English translation: NACA TM 1261 (1950)}
}

@article{Zeldovich1970,
  author  = {Zel'dovich, Ya. B. and Librovich, V. B. and Makhviladze, G. M. and Sivashinsky, G. I.},
  title   = {On the development of detonation in a non-uniformly preheated gas},
  journal = {J. Appl. Mech. Tech. Phys.},
  volume  = {11},
  number  = {2},
  pages   = {264--270},
  year    = {1970}
}

@article{HeClavin1994,
  author  = {He, L. and Clavin, P.},
  title   = {On the direct initiation of gaseous detonations by an energy source},
  journal = {J. Fluid Mech.},
  volume  = {277},
  pages   = {227--248},
  year    = {1994}
}

@book{LandauLifshitz1987,
  author    = {Landau, L. D. and Lifshitz, E. M.},
  title     = {Fluid Mechanics},
  edition   = {2nd},
  series    = {Course of Theoretical Physics},
  volume    = {6},
  publisher = {Pergamon Press},
  address   = {Oxford},
  year      = {1987}
}

@book{Williams1985,
  author    = {Williams, F. A.},
  title     = {Combustion Theory},
  edition   = {2nd},
  publisher = {Benjamin/Cummings},
  address   = {Menlo Park, CA},
  year      = {1985}
}

@book{ClavinSearby2016,
  author    = {Clavin, P. and Searby, G.},
  title     = {Combustion Waves and Fronts in Flows},
  publisher = {Cambridge University Press},
  address   = {Cambridge},
  year      = {2016}
}

@article{Zeldovich1948,
  author  = {Zel'dovich, Ya. B.},
  title   = {On the theory of flame propagation},
  journal = {Zh. Fiz. Khim.},
  volume  = {22},
  pages   = {27--49},
  year    = {1948},
  note    = {English translation: NACA TM 1282 (1951)}
}

@techreport{Linan1971,
  author      = {Li\~{n}\'{a}n, A.},
  title       = {A theoretical analysis of premixed flame propagation with an isothermal chain reaction},
  institution = {Instituto Nacional de T\'{e}cnica Aeroespacial},
  address     = {Madrid},
  type        = {Technical Report},
  number      = {AFOSR Contract E00AR68-0031, No.~1},
  year        = {1971}
}

@article{Dold2002,
  author  = {Dold, J. W. and Thatcher, R. W. and Omon-Arancibia, A. and Redman, J.},
  title   = {From one-step to chain-branching premixed-flame asymptotics},
  journal = {Proc. Combust. Inst.},
  volume  = {29},
  pages   = {1519--1526},
  year    = {2002}
}

@article{Dold2007,
  author  = {Dold, J. W.},
  title   = {Premixed flames modelled with thermally sensitive intermediate branching kinetics},
  journal = {Combust. Theory Model.},
  volume  = {11},
  pages   = {909--948},
  year    = {2007}
}

@article{LeeKim2024,
  author  = {Lee, S. R. and Kim, J. S.},
  title   = {The asymptotic solution of near-limit chain-branching premixed flames with the {Zel'dovich--Li\~{n}\'{a}n} two-step mechanism in the linear and fast recombination regime},
  journal = {Combust. Flame},
  volume  = {265},
  pages   = {113441},
  year    = {2024}
}

@article{Joulin1985,
  author  = {Joulin, G. and Li\~{n}\'{a}n, A. and Ludford, G. S. S. and Peters, N. and Schmidt-Lain\'{e}, C.},
  title   = {Flames with chain-branching/chain-breaking kinetics},
  journal = {SIAM J. Appl. Math.},
  volume  = {45},
  pages   = {420--434},
  year    = {1985}
}

@article{SanchezWilliams2014,
  author  = {S\'{a}nchez, A. L. and Williams, F. A.},
  title   = {Recent advances in understanding of flammability characteristics of hydrogen},
  journal = {Prog. Energy Combust. Sci.},
  volume  = {41},
  pages   = {1--55},
  year    = {2014}
}

@article{ShortQuirk1997,
  author  = {Short, M. and Quirk, J. J.},
  title   = {On the nonlinear stability and detonability limit of a detonation wave for a model three-step chain-branching reaction},
  journal = {J. Fluid Mech.},
  volume  = {339},
  pages   = {89--119},
  year    = {1997}
}

@article{ShortSharpe2003,
  author  = {Short, M. and Sharpe, G. J.},
  title   = {Pulsating instability of detonations with a two-step chain-branching reaction model: theory and numerics},
  journal = {Combust. Theory Model.},
  volume  = {7},
  pages   = {401--416},
  year    = {2003}
}

@article{SharpeMaflahi2006,
  author  = {Sharpe, G. J. and Maflahi, N.},
  title   = {Homogeneous explosion and shock initiation for a three-step chain-branching reaction model},
  journal = {J. Fluid Mech.},
  volume  = {566},
  pages   = {163--194},
  year    = {2006}
}

@article{Sanchez2001,
  author  = {S\'{a}nchez, A. L. and Carretero, M. and Clavin, P. and Williams, F. A.},
  title   = {One-dimensional overdriven detonations with branched-chain kinetics},
  journal = {Phys. Fluids},
  volume  = {13},
  pages   = {776--792},
  year    = {2001}
}

@article{Ng2005,
  author  = {Ng, H. D. and Radulescu, M. I. and Higgins, A. J. and Nikiforakis, N. and Lee, J. H. S.},
  title   = {Numerical investigation of the instability for one-dimensional {Chapman--Jouguet} detonations with chain-branching kinetics},
  journal = {Combust. Theory Model.},
  volume  = {9},
  pages   = {385--401},
  year    = {2005}
}

@article{LeeStewart1990,
  author  = {Lee, H. I. and Stewart, D. S.},
  title   = {Calculation of linear detonation instability: one-dimensional instability of plane detonation},
  journal = {J. Fluid Mech.},
  volume  = {216},
  pages   = {103--132},
  year    = {1990}
}

@article{Erpenbeck1964,
  author  = {Erpenbeck, J. J.},
  title   = {Stability of idealized one-reaction detonations},
  journal = {Phys. Fluids},
  volume  = {7},
  pages   = {684--696},
  year    = {1964}
}

@article{KasimovStewart2004,
  author  = {Kasimov, A. R. and Stewart, D. S.},
  title   = {On the dynamics of self-sustained one-dimensional detonations: a numerical study in the shock-attached frame},
  journal = {Phys. Fluids},
  volume  = {16},
  pages   = {3566--3578},
  year    = {2004}
}

@article{Henrick2006,
  author  = {Henrick, A. K. and Aslam, T. D. and Powers, J. M.},
  title   = {Simulations of pulsating one-dimensional detonations with true fifth order accuracy},
  journal = {J. Comput. Phys.},
  volume  = {213},
  pages   = {311--329},
  year    = {2006}
}

@article{Taylor1950,
  author  = {Taylor, G. I.},
  title   = {The dynamics of the combustion products behind plane and spherical detonation fronts in explosives},
  journal = {Proc. R. Soc. Lond. A},
  volume  = {200},
  pages   = {235--247},
  year    = {1950}
}

@book{Toro2009,
  author    = {Toro, E. F.},
  title     = {Riemann Solvers and Numerical Methods for Fluid Dynamics},
  edition   = {3rd},
  publisher = {Springer},
  address   = {Berlin},
  year      = {2009}
}

@article{JiangShu1996,
  author  = {Jiang, G.-S. and Shu, C.-W.},
  title   = {Efficient implementation of weighted {ENO} schemes},
  journal = {J. Comput. Phys.},
  volume  = {126},
  pages   = {202--228},
  year    = {1996}
}

@article{ShuOsher1988,
  author  = {Shu, C.-W. and Osher, S.},
  title   = {Efficient implementation of essentially non-oscillatory shock-capturing schemes},
  journal = {J. Comput. Phys.},
  volume  = {77},
  pages   = {439--471},
  year    = {1988}
}

@article{Kessler2012,
  author  = {Kessler, D. A. and Gamezo, V. N. and Oran, E. S.},
  title   = {Gas-phase detonation propagation in mixture composition gradients},
  journal = {Phil. Trans. R. Soc. A},
  volume  = {370},
  pages   = {567--596},
  year    = {2012}
}

@article{IshiiKojima2007,
  author  = {Ishii, K. and Kojima, M.},
  title   = {Behavior of detonation propagation in mixtures with concentration gradients},
  journal = {Shock Waves},
  volume  = {17},
  pages   = {95--102},
  year    = {2007}
}

@book{LinanWilliams1993,
  author    = {Li\~{n}\'{a}n, A. and Williams, F. A.},
  title     = {Fundamental Aspects of Combustion},
  publisher = {Oxford University Press},
  address   = {New York},
  year      = {1993}
}

@article{ClavinHe1996,
  author  = {Clavin, P. and He, L.},
  title   = {Stability and nonlinear dynamics of one-dimensional overdriven detonations in gases},
  journal = {J. Fluid Mech.},
  volume  = {306},
  pages   = {353--378},
  year    = {1996}
}

@article{ClavinWilliams2002,
  author  = {Clavin, P. and Williams, F. A.},
  title   = {Dynamics of planar gaseous detonations near {Chapman--Jouguet} conditions for small heat release},
  journal = {Combust. Theory Model.},
  volume  = {6},
  pages   = {127--139},
  year    = {2002}
}

@article{ClavinWilliams2009,
  author  = {Clavin, P. and Williams, F. A.},
  title   = {Multidimensional stability analysis of gaseous detonations near {Chapman--Jouguet} conditions for small heat release},
  journal = {J. Fluid Mech.},
  volume  = {624},
  pages   = {125--150},
  year    = {2009}
}

@article{ClavinWilliams2012,
  author  = {Clavin, P. and Williams, F. A.},
  title   = {Analytical studies of the dynamics of gaseous detonations},
  journal = {Phil. Trans. R. Soc. A},
  volume  = {370},
  pages   = {597--624},
  year    = {2012}
}

@article{Clavin2004,
  author  = {Clavin, P.},
  title   = {Theory of gaseous detonations},
  journal = {Chaos},
  volume  = {14},
  pages   = {825},
  year    = {2004}
}

@article{ClavinDenet2018,
  author  = {Clavin, P. and Denet, B.},
  title   = {Decay of plane detonation waves to the self-propagating {Chapman--Jouguet} regime},
  journal = {J. Fluid Mech.},
  volume  = {845},
  pages   = {170--202},
  year    = {2018}
}

@article{Clavin2019,
  author  = {Clavin, P.},
  title   = {Advances in the analytical study of the dynamics of gaseous detonation waves},
  journal = {C. R. M\'{e}canique},
  volume  = {347},
  pages   = {273--286},
  year    = {2019}
}

@article{ClavinDenet2020,
  author  = {Clavin, P. and Denet, B.},
  title   = {Analytical study of the direct initiation of gaseous detonations for small heat release},
  journal = {J. Fluid Mech.},
  volume  = {897},
  pages   = {A30},
  year    = {2020}
}

@article{Clavin2021,
  author  = {Clavin, P. and Hern\'{a}ndez S\'{a}nchez, R. and Denet, B.},
  title   = {Asymptotic analysis of the critical dynamics of spherical gaseous detonations},
  journal = {J. Fluid Mech.},
  volume  = {915},
  pages   = {A122},
  year    = {2021}
}

@techreport{Chandrasekhar1943,
  author      = {Chandrasekhar, S.},
  title       = {On the decay of plane shock waves},
  institution = {Ballistic Research Laboratory, Aberdeen Proving Ground},
  type        = {Report},
  number      = {423},
  year        = {1943}
}

@incollection{Peters1985,
  author    = {Peters, N.},
  title     = {Numerical and asymptotic analysis of systematically reduced reaction schemes for hydrocarbon flames},
  booktitle = {Numerical Simulation of Combustion Phenomena},
  editor    = {Glowinski, R. and Larrouturou, B. and Temam, R.},
  series    = {Lecture Notes in Physics},
  volume    = {241},
  pages     = {90--109},
  publisher = {Springer},
  address   = {Berlin},
  year      = {1985}
}

@book{ZeldovichRaizer1966,
  author    = {Zel'dovich, Ya. B. and Raizer, Yu. P.},
  title     = {Physics of Shock Waves and High-Temperature Hydrodynamic Phenomena},
  publisher = {Academic Press},
  address   = {New York},
  year      = {1966},
  note      = {Reprinted by Dover, Mineola, NY, 2002}
}

@article{PetersWilliams1987,
  author  = {Peters, N. and Williams, F. A.},
  title   = {The asymptotic structure of stoichiometric methane--air flames},
  journal = {Combust. Flame},
  volume  = {68},
  pages   = {185--207},
  year    = {1987}
}

@article{Friedrichs1948,
  author  = {Friedrichs, K. O.},
  title   = {Formation and decay of shock waves},
  journal = {Commun. Pure Appl. Math.},
  volume  = {1},
  pages   = {211--245},
  year    = {1948}
}

@article{Napieralski2027,
  author  = {Napieralski, M. and Huete, C. and Coenen, W. and Kurdyumov, V. N. and Douglas, C. M. and S\'{a}nchez-Sanz, M.},
  title   = {Modal, non-modal and sensitivity analysis of planar deflagrations with low molecular diffusivity and branching kinetics},
  journal = {Appl. Math. Model.},
  volume  = {161},
  pages   = {117152},
  year    = {2027}
}

@article{Carmona2026,
  author  = {Carmona, R. and Fern\'{a}ndez-Galisteo, D. and Jim\'{e}nez, C. and Kurdyumov, V. N.},
  title   = {A new propagation regime below the flammability limit for flames governed by two-step chain-branching kinetics: numerical solutions and asymptotic analysis},
  journal = {Combust. Flame},
  volume  = {289},
  pages   = {115042},
  year    = {2026}
}

@article{Cuadra2020,
  author  = {Cuadra, A. and Huete, C. and Vera, M.},
  title   = {Effect of equivalence ratio fluctuations on planar detonation discontinuities},
  journal = {J. Fluid Mech.},
  volume  = {903},
  pages   = {A30},
  year    = {2020}
}

@article{Huete2013,
  author  = {Huete, C. and S\'{a}nchez, A. L. and Williams, F. A.},
  title   = {Theory of interactions of thin strong detonations with turbulent gases},
  journal = {Phys. Fluids},
  volume  = {25},
  number  = {7},
  year    = {2013}
}

@article{Huete2014,
  author  = {Huete, C. and S\'{a}nchez, A. L. and Williams, F. A.},
  title   = {Linear theory for the interaction of small-scale turbulence with overdriven detonations},
  journal = {Phys. Fluids},
  volume  = {26},
  pages   = {116101},
  year    = {2014}
}

@article{MartinezRuiz2020,
  author  = {Mart\'{i}nez-Ruiz, D. and Huete, C. and S\'{a}nchez, A. L. and Williams, F. A.},
   title   = {Theory of weakly exothermic oblique detonations},
  journal = {AIAA J.},
   volume  = {58},
   number  = {1},
   pages   = {236--242},
   year    = {2020}
}

@article{fernandez2019one,
  title={A one-step reduced mechanism for near-limit hydrogen combustion with general stoichiometry},
  author={Fern{\'a}ndez-Galisteo, D. and Weiss, A. D. and S{\'a}nchez, A. L. and Williams, F. A.},
  journal={Combust. Flame},
  volume={208},
  pages={1--4},
  year={2019},
  publisher={Elsevier}
}

@article{carpio2020near,
  title={Near-limit {H}2-{O}2-{N}2 combustion in nonpremixed counterflow mixing layers},
  author={Carpio, J. and Rajamanickam, P. and S{\'a}nchez, A. L. and Williams, F. A.},
  journal={Combust. Flame},
  volume={216},
  pages={426--438},
  year={2020},
  publisher={Elsevier}
}

\end{document}